**Femtosecond Optical Two-Way Time-Frequency Transfer in the Presence of Motion**

Laura C. Sinclair, Hugo Bergeron, William C. Swann, Isaac Khader, Kevin C. Cossel, Michael Cermak, Nathan R. Newbury, and Jean-Daniel Deschênes

[1] *National Institute of Standards and Technology, 325 Broadway, Boulder, Colorado 80305*

[2] *Université Laval, 2325 Rue de l'Université, Québec, QC, G1V 0A6, Canada*

Platform motion poses significant challenges to high-precision optical time and frequency transfer. We give a detailed description of these challenges and their solutions in comb-based optical two-way time and frequency transfer (O-TWTFT). Specifically, we discuss the breakdown in reciprocity due to relativity and due to asynchronous sampling, the impact of optical and electrical dispersion, and velocity-dependent transceiver calibration. We present a detailed derivation of the equations governing comb-based O-TWTFT in the presence of motion. We describe the implementation of real-time signal processing algorithms based on these equations and demonstrate active synchronization of two sites over turbulent air paths to below a femtosecond time deviation despite effective velocities of ±25 m/s, which is the maximum achievable with our physical setup. With the implementation of the time transfer equation derived here, we find no velocity-dependent bias between the synchronized clocks to within at two-sigma statistical uncertainty of 330 attoseconds.

# I. Introduction

Frequency-comb-based optical two-way time-frequency transfer (O-TWTFT) has progressed rapidly in the last few years from a straightforward frequency comparison using a large lab system [1], to real-time synchronization with a potentially fieldable system [2], to generation of coherent microwaves at remote sites [3], and to operation over a strongly turbulent 12-km path [4]. Nevertheless, compared to mature, deployed fiber-based approaches [5–15], O-TWTFT is still at an early stage. Indeed, Refs. [1–4] all demonstrate operation over links with a slowly varying time-of-flight, whose fluctuations are exclusively due to turbulence or minute platform vibrations. . However, free-space networks will have to cope with motion between the clock sites, which is a complication avoided in fiber-optic networks. At motion of 30 m/s, e.g that of a car driving on a highway, these systems would suffer errors in the tens of picoseconds -- a performance degradation of ten thousand or more. Multiple velocity-dependent effects, some fundamental and some implementation specific, cause these errors, and must be understood and accounted for at the femtosecond level. In order to return to femtosecond-level synchronization despite significant motion, a new implementation of O-TFTFT is required where all the available information is used to address the various effects.

Here, we derive the basic equations for comb-based O-TWTFT that compensates for motion. We present our hardware implementation and demonstrate femtosecond time synchronization between two clock sites. Since a mobile optical clock was unavailable, we introduce motion by rapidly changing the path length between clocks. (The velocity-dependent effects are similar to a moving clock with the exception of time dilation.) We find no velocity-dependent degradation of time synchronization to within a two-sigma uncertainty of 330 attoseconds, correspondingly, no degradation of frequency syntonization down to $2\text{x}10^{-18}$ in fractional frequency uncertainty (modified Allan deviation). This paper is closely related to the work summarized in Ref. [16] and serves to more closely examine the effects of motion and their solutions and analyze the possibility of residual bias.

The paper is organized as follows. Section II provides an overview of the comb-based O-TWTFT setup. Section III begins with an overview of velocity-dependent effects on clock synchronization before deriving the complete set of equations to compute the clock offset between the sites without systematic error in the presence of motion. Section IV describes the hardware implementation. Section V presents results. Section VI discusses scaling to higher velocities than those achievable with our experimental testbeds, and finally Section VII concludes. We consider here only the effects of closing velocity between the two sites. There are potential turbulence-related systematics due to transverse velocity between the two sites, but these are expected to be minimal [17,18] as borne out by a recent experiment [19].

## II. Overview of the experimental setup and comb-based O-TWTFT

Our goal is to compare the time (and therefore frequency) between two remote clocks, located at site A and B. Furthermore, we implement that time comparison in real-time such that the results can be used to actively synchronize, or phase lock, the clock at site B to the one at site A. This paper discusses the full system required for such time synchronization between clocks in the presence of motion. It is worth emphasizing that a frequency comparison between sites is considerably simpler and may be sufficient for some applications. Its implementation represents a subset of the system described here.

In order to verify that we have achieved synchronization at the femtosecond-level in time (and $10^{-18}$ in fractional frequency), we use a folded link as in previous work to allow for a direct out-of-loop time comparison between sites [2]. We emphasize that all communication and timing signals associated with the O-TWTFT traverse only the long free-space link.

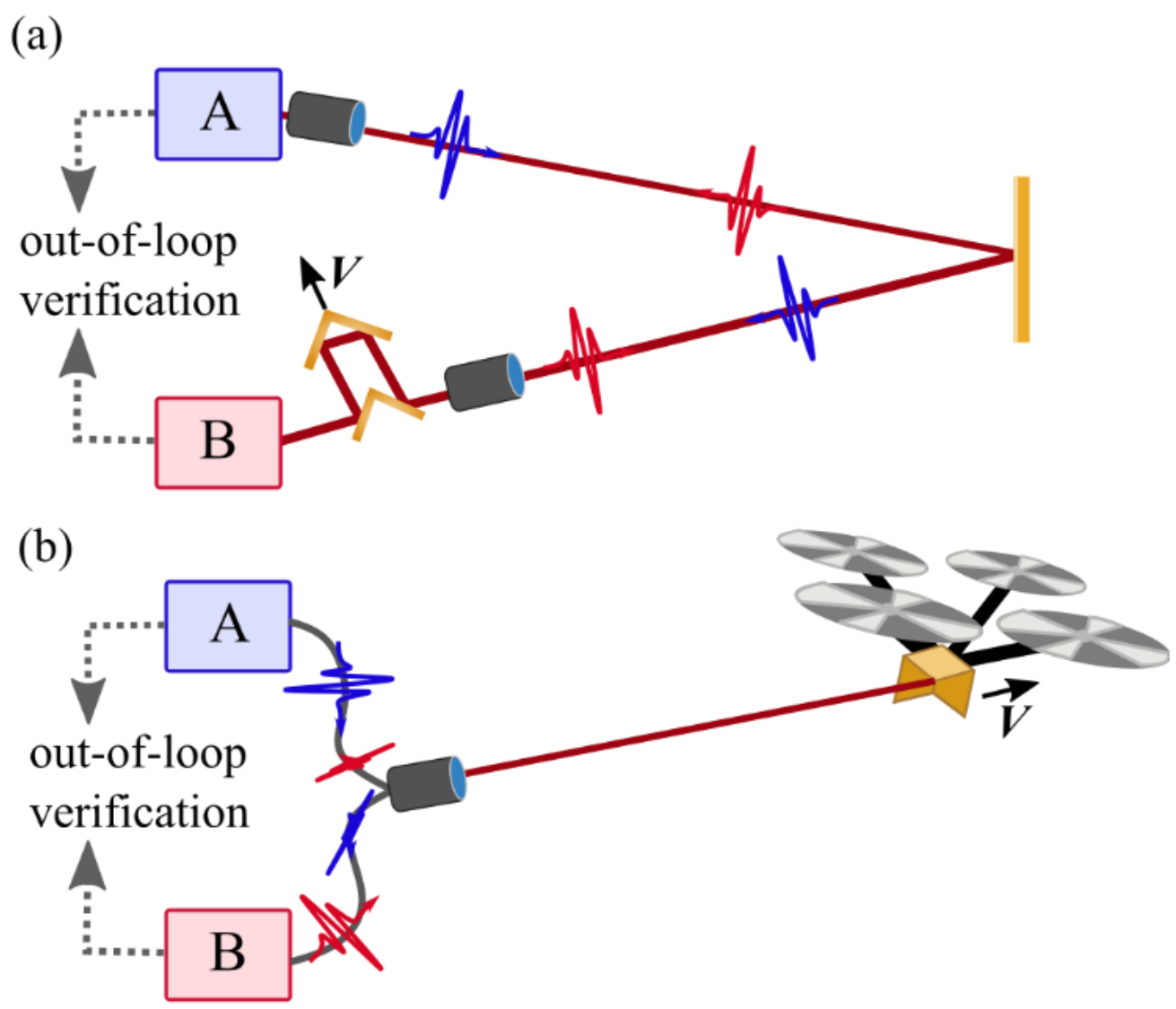

Figure 1: Experimental setup to evaluate comb-based O-TWTFT with motion. (a) A multi-passed moving retroreflector is located adjacent to clock B to simulate clock B motion with effective closing velocity, *V*. (b) A quadcopter-mounted-retroreflector is flown approximately equidistant from sites A and B.

Figure 1 illustrates the two configurations used here to implement and test O-TWTFT with a rapidly changing path length between Sites A and B. We insert a moving retroreflector in the path either mounted on a traveling rail or on a quadcopter. The latter geometry mimics that expected for an intermediate passive air-borne platform connecting two fixed sites, except for motion transverse to the link. (In separate work, we find the effect of this transverse motion should be observable but is still below a few femtoseconds [17–19].)

The basic configuration of the comb-based O-TWTFT is shown in Fig. 2 and follows Ref. [2,3,20]. At each site, there is a clock (Fig. 2a), an optical transceiver for the comb-based timing, an optical transceiver for the coherent optical communication channel, and a real-time digital signal processing system comprised of a field programmable gate array (FPGA) and digital signal processor (DSP) platform. As shown in Fig. 2a and as in Ref. [2], we construct our clock by phase-locking a frequency comb with repetition frequency, $f_r$ ~ 200 MHz, to an ~ 195-THz optical oscillator. The time is then defined by the arrival of the labelled comb pulses at a given reference plane. Here, we define site A as the master site and site B as the remote site, at which we apply feedback to synchronize it to the master site.

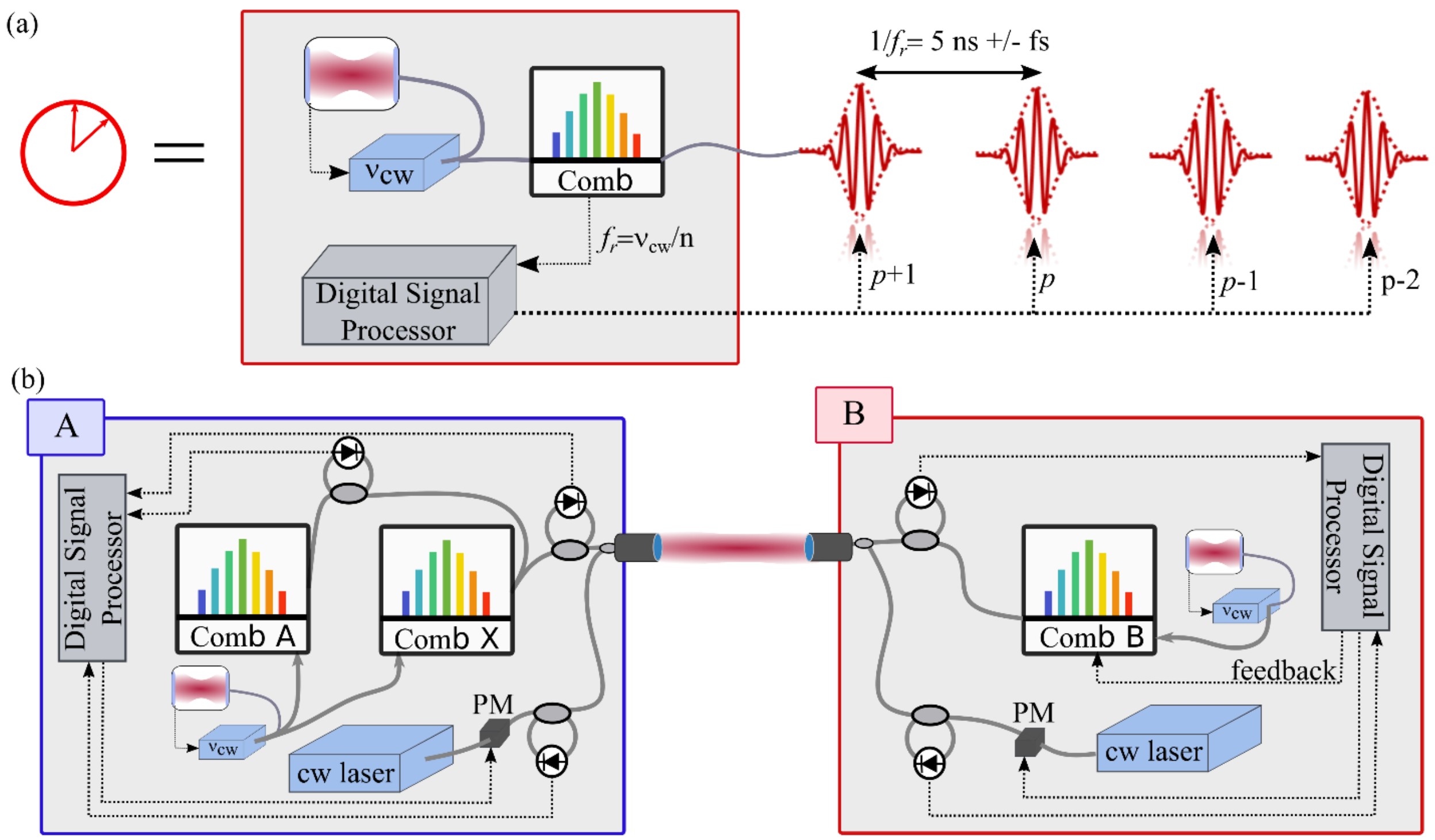

Figure 2: (a) Definition of the "clock" or timescale at each site. A self-referenced frequency comb is phase-locked to a cavity-stabilized laser at $\nu_{cw}$ ~ 195 THz. The comb produces a phase coherent pulse train with repetition period, $1/f_r \sim 5$ ns and femtosecond pulse-to-pulse timing jitter. A digital signal processor enables the labelling of each pulse, corresponding to the "ticks" of the clock. (b) Basic comb-based O-TWTFT configuration, as described in the text. PM: phase modulator

As shown in Figure 2, at each site A and B we place a "clock" comb, A and B respectively, that form the time base. To accomplish the time-transfer, we also introduce a transfer comb X at site A with a repetition frequency offset by $\Delta f_r$ ~ 2 kHz. Timing information is then exchanged between sites via the two-way exchange of light from this comb X and comb B. We achieve femtosecond-level detection of the arrival times of the transmitted pulses versus the local comb at each site through linear optical sampling [2,21], or, in other words, by measuring the heterodyne signal between (i) the incoming remote comb B and transfer comb X at the master site A, and (ii) the incoming transfer comb X and remote comb B at the remote site B. In addition, we measure (iii) the heterodyne signal between the master comb A and transfer comb X at the master site A to establish their relative timing. When combined, these three heterodyne comb signals provide relative time information between the two sites at the femtosecond level, if properly interpreted, but suffer from an ambiguity of ~ 5 ns (the separation of the comb pulses).

To remove this ambiguity, we operate an optical-communication-based TWTFT in parallel. We establish a coherent, single-mode optical communication link between sites by wavelength multiplexing a phase-modulated cw laser with the comb light and transmitting it via the same free-space optical terminal. This optical communication channel operates at 10 Mbps with Manchester coding and is described in detail in Ref. [22]. It serves two purposes. First, we use it to implement a communication-based TWTFT that provides the "coarse" time offset between sites with < 100 ps precision, which is more than sufficient to remove the 5-ns ambiguity from the heterodyne comb signals. Second, we use it to transmit the timing information recorded at the master site A to site B.

Once the timing data are collected at the remote site B, which occurs every $1/\Delta f_r \sim 0.5$ ms, the data are processed in a digital signal processor, that implements the equations derived in the next section to compute the clock offset. When actively synchronized, this clock offset is fed into a Kalman-filter whose output feeds a proportional integral controller to adjust the timing of the comb B pulse train (i.e. the clock at site B) for zero clock offset.

We note that, for an experiment that requires only frequency comparison in post processing, much of this hardware is not required including: comb X, the coherent communication channel and associated communication-based OTWTF, the real-time digital signal processor, and the extensive transceiver calibration. In that case, it is sufficient to simply exchange pulses from two offset combs, record their heterodyne signals, and process them offline. In implementing comb-based O-TWTFT, it is therefore critical to identify the requirements of the overall system in terms of frequency or time and thus the minimum required setup.

# III. Derivation of the Timing Equations with Motion

In this section, we first discuss two-way time transfer in terms of the exchange of a pair of directly-detected pulses for several reasons. First, it is a useful starting point for the more complex actual system. Second, the communication-based O-TWTFT follows this basic prescription fairly closely. Third, it is relatively straightforward to understand the five basic velocity-dependent effects in this standard picture before addressing them in the context of comb-based O-TWTFT.

## III.A Overview of Main Velocity-Dependent Effects

First, consider the two-way exchange of an optical pulse pair between two fixed sites. In the simplest picture, we measure the departure time, $T_{AA}$, of the pulse from site A against the site A timebase and the arrival time, $T_{AB}$, of the same pulse at site B against the site B timebase. It must be that $T_{AB} = T_{AA} + T_{A\to B} - \Delta t_{AB}$, where $\Delta t_{AB}$ is the slowly-varying clock offset between site A and B's timebases and $T_{A\to B} = L/c$ is the time-of-flight from A to B with $L$ the potentially time-varying path length and $c$ the speed of light across the path. We similarly send a pulse from site B to site A to generate $T_{BB}$, and $T_{BA} = T_{BB} + T_{B\to A} + \Delta t_{AB}$ where $T_{B\to A} = L/c$ is the time-of-flight from B to A. For a constant distance $L$, a simple linear combination provides the well-known basic two-way formula [23–25],

$$\Delta t_{AB} = \frac{1}{2}\left[T_{AA} - T_{AB} - T_{BB} + T_{BA}\right] + \frac{1}{2}\left[T_{A\to B} - T_{B\to A}\right] + \Delta T_{cal} \quad (1)$$

where $\Delta T_{cal}$ is some calibration constant for time offsets in the transceivers. For a fully reciprocal link, i.e. $T_{A\to B} \equiv T_{B\to A}$, the second term is exactly zero and the explicit time-of-flight dependence vanishes. However, if the time-of-flight difference, $\left[T_{A\to B} - T_{B\to A}\right]$, is non-zero, i.e. non-reciprocal, but the link is falsely assumed to be reciprocal, this non-reciprocity introduces a systematic error in $\Delta t_{AB}$. With motion, such non-reciprocity arises from two effects: asynchronous sampling of the two-way signals and relativistic non-reciprocity.

In addition to these two effects, motion will lead to errors in $\Delta t_{AB}$ for three other reasons. First, the Doppler shifts on the received signals can introduce a systematic velocity dependence in the measured arrival times, $T_{AB}$ and $T_{BA}$. Second, $\Delta T_{cal}$ is velocity dependent as the multiple transceiver paths carry optical signals which are both unshifted, if local, or Doppler shifted, if remote. Third, motion complicates the resolution of timing ambiguities associated with the use of periodic waveforms, such as the pulses of a frequency comb. Table I quantifies these different effects that are described in more detail below.

III.A.1 *Fundamental Relativistic Breakdown in Reciprocity*

First consider the scenario in Fig. 3a where each site emits its pulse at exactly the same time in site A's reference frame while site B is moving away from site A. Despite the simultaneous emission time, the time-of-flight for the two directions is not equal, i.e. $T_{A\to B} \neq T_{B\to A}$, and their difference is $T_{A\to B} - T_{B\to A} = LV/c^2$ to first order in *V*, the closing velocity of site B with respect to the fixed site A, where *L* is the instantaneous clock separation. For modest values of *L* = 4 km and $V$ = 25 m/s, the non-reciprocal time-of-flight is 1 ps, which would cause a corresponding 1-ps error in $\Delta t_{AB}$ if uncorrected. However, with an appropriate velocity estimate, we can correctly include the non-reciprocal time-of-flight in Eqn. (1).

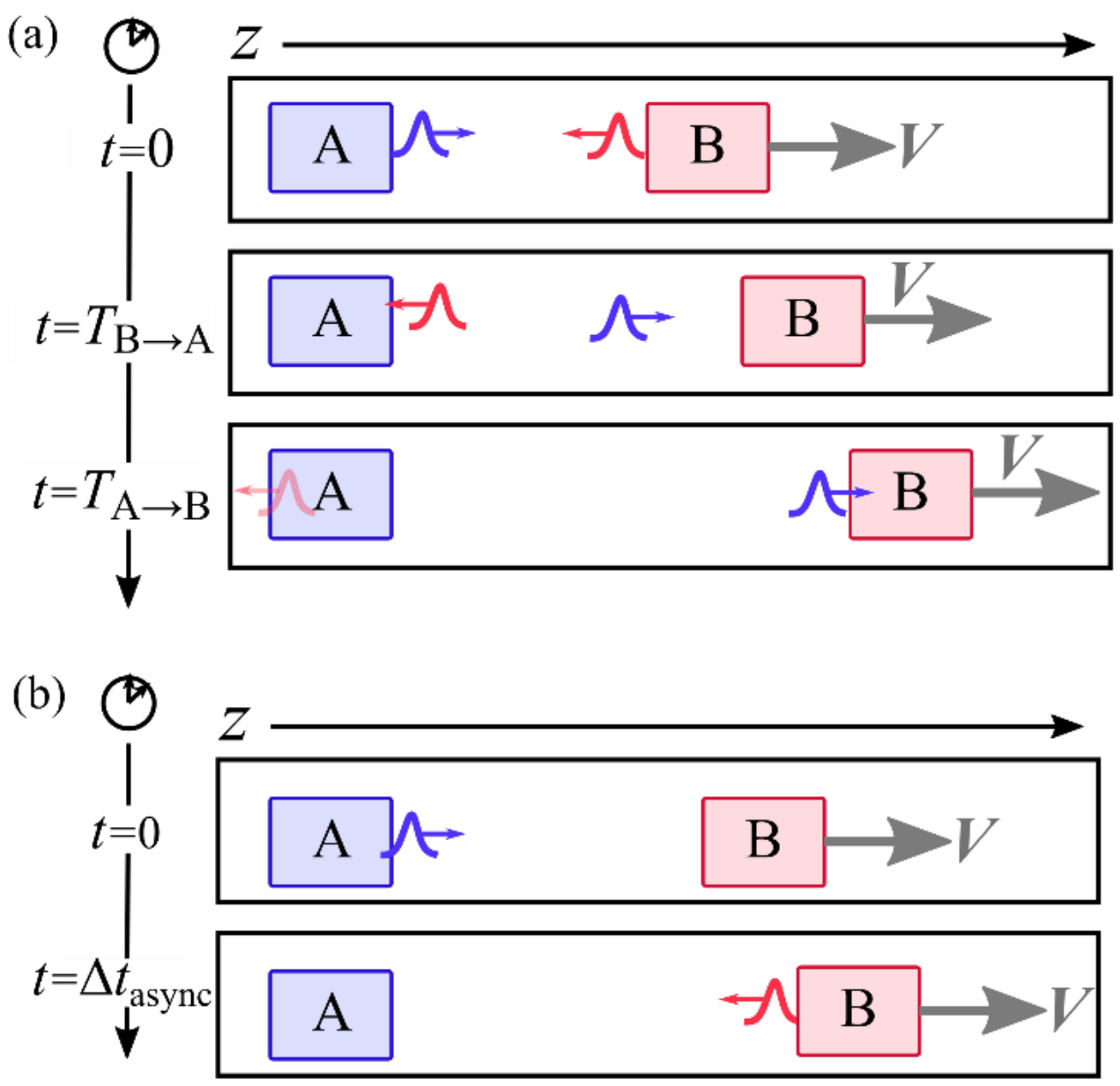

Figure 3: (a) Fundamental relativistic breakdown in reciprocity. Pulses launched simultaneously from static clock A and moving clock B in site A's reference frame experience a non-reciprocal time-of-flight with $T_{A\to B} \neq T_{B\to A}$. (b) Breakdown in reciprocity due to asynchronous sampling. Pulses launched at different times (asynchronously) from static clock A and moving clock B (again relative to Site A's reference frame), e.g. at $t=0$ and at $t=\Delta t_{async}$, experience different time-of-flights.

III.A.2 *Breakdown in Reciprocity due to Asynchronous Sampling*

Now consider the scenario in Fig. 3b, where the timing signals from the two sites are launched asynchronously with a time offset $\Delta t_{async} = T_{AB} - T_{BA} + \Delta t_{AB}$ (as will invariably be the case in any real system). With motion, this leads to a non-reciprocal time-of-flight of $T_{A\to B} - T_{B\to A} = \Delta t_{async} V / c$. For O-TWTFT, the use of a transfer comb and linear optical sampling all but guarantees asynchronous sampling with $\Delta t_{async}$ ranging from 0 to $1/(2\Delta f_r)$. At $V$ = 25 m/s and $\Delta f_r = 2$ kHz, the non-reciprocal time-of-flight is 20 ps, which would cause a corresponding 20-ps error in $\Delta t_{\mathrm{AB}}$ if uncorrected. However, this error can be avoided by interpolation of the timing signals to a common measurement time or equivalently by including a correction factor in (1).

III.A.3 *Delay-Doppler Coupling*

With motion, the light pulses suffer Doppler shifts. When combined with the system dispersion, this leads to a delay-Doppler coupling, which amounts to a systematic error in the measured arrival times $T_{\mathrm{AB}}$ and $T_{\mathrm{BA}}$. For our optical heterodyne system, there are three distinct effects.

First, for an optical link with lumped dispersion, $\beta_2^{path}$, the systematic error in arrival time is $2\pi\nu_c\beta_2^{path}V/c$, where $\nu_c$ ~ 200 THz is the optical carrier. For our system, $\beta_2^{path}$ is between 0.1 and 1 $\mathrm{ps}^2$, due to fiber-optic leads, telescope optics, and the air path. This effect is then of order 10 - 100 attoseconds for $V$ = 25 m/s.

Second, we use linear optical sampling between frequency comb pulse trains to achieve femtosecond precision in the measurement of the pulse arrival times. In linear optical sampling, we heterodyne the incoming pulse train of repetition rate, $f_r$ ~ 200 MHz, against a local pulse train with an offset repetition rate, $f_r \pm \Delta f_r$. The resulting heterodyne signal is a series of pulse bursts, or interferograms, in the rf domain that repeat at $\Delta f_r$ ~ 2 kHz. The advantage of this approach is that any time shift in the incoming pulse train is amplified by a factor of $f_r/\Delta f_r \sim 10^5$ in the timing of the rf-domain interferograms, thus enabling femtosecond timing precision. However, there is a penalty associated with this linear optical sampling -- the timing error from the product of any Doppler shift and differential chirp, $\Delta\beta_2^{Combs}$, between the heterodyned comb pulses is amplified by $f_r/\Delta f_r$. This delay-Doppler systematic is thus of order $\left(f_r/\Delta f_r\right)2\pi\nu_c\Delta\beta_2^{Combs}V/c$. (See

Appendix A for a derivation.) For our parameters, the resulting systematic error can exceed 1 ps without compensation,.

To suppress these optical-dispersion effects to well below 1 fs, a simple linear correction based on the above equations and an estimate of $\Delta\beta_2^{Combs}$ and $\beta_2^{path}$ is insufficient. Instead, we use a two-pronged approach. We first reduce the optical dispersion, $\Delta\beta_2^{Combs}$ and $\beta_2^{path}$, through the addition of dispersion-compensating fibers before the optical telescopes. Second, we adopt a technique from the RADAR community and find the effective arrival time of the received pulse from the peak of the cross-ambiguity function between the measured interferogram and expected signal. The cross-ambiguity function search efficiently removes any remaining dispersion to all orders, meaning that the hardware dispersion compensation does not have to be exact.

The third delay-Doppler systematic occurs in the rf domain. The interferograms generated by the linear optical sampling are an rf pulse, described by an rf carrier and envelope. Any optical Doppler shift is mapped directly to the rf carrier, leading to a timing error in the rf domain of the interferograms of $2\pi\nu_c\beta_2^{RF}V/c$, where $\beta_2^{RF}$ is the RF dispersion from photodetector responses, electrical filters, impedance mismatches etc. This timing error can reach nanoseconds for large (10's of MHz) Doppler shifts. However, as mentioned above, the optical pulse arrival time is found by dividing the interferogram arrival time in the rf domain by a factor of $f_r/\Delta f_r$ ~ $10^5$, which greatly suppresses any error. This is the inverse of the amplification factor described above. Nevertheless, we must apply a compensation filter, calculated during the system calibration, to the digitized rf signals to effectively set $\beta_2^{RF}$ close to zero and therefore achieve sub-femtosecond timing.

III.A.4 *Velocity-Dependent Transceiver Calibration*

In the simplest case, the calibration constant, $\Delta T_{\text{cal}}$, in Eq. (1) reflects a time delay in the transceiver between the reference plane and the detection of the incoming pulses. However, each transceiver is far from a compact point and consists of a distributed set of optical components comprising optical oscillators, frequency combs, modulated cw lasers, optical transceivers for detecting the arrival time of frequency comb pulses, and optical transceivers for the communication-based O-TWTFT as is illustrated in Fig. 2 (and later in Fig. 4). Nevertheless, in the absence of significant Doppler shifts, as in Refs. [2–4], the calibration of this distributed system can still be lumped into a single overall time offset, $\Delta T_{\text{cal}}$. Here, with Doppler shifts, that is no longer the case and this calibration must be expanded to include a velocity-dependent contribution, $\Delta T_{\text{cal}} \rightarrow \Delta T_{\text{cal}} + (V/c)\Delta T_{\text{cal}}^{V}$. The computation of these calibration terms requires in-depth probing of the various delays in the transceiver via an rf-domain optical time domain reflectometer (OTDR). Moreover, there are additional calibration terms associated with the use of a Doppler simulator rather than a moving clock. (See Section III.B.4 for a detailed description.) As indicated in Table I, the maximum error is estimated based on the total transceiver path length, $L_{transceiver}$, as $VL_{\text{transceiver}}/c^2$.

III.A.5 *Periodic Waveform Ambiguities*

In our LOS approach, there is an integer ambiguity associated with the relative pulse number of the two interfering pulse trains – i.e. exactly which pulses overlapped at the transceiver. We ultimately measure three such interferograms, each with its own integer ambiguity, and combine their timing to compute the overall clock offset. For the fixed terminal case, we can combine these three integer ambiguities into a single integer, which is then resolved from the communication-channel O-TWTFT. With motion however, these integers enter the overall clock offset computation with different scale factors. (See Eqn. (20).) Moreover, the combination of motion and turbulence-induced signal fades means that the time-of-flight could change by more than the 5-ns ambiguity range between measurements. To counter these problems, the timing data from the communication-based O-TWTFT and comb-based O-TWTFT is combined early in the processing such that any ambiguities are resolved independently for each interferogram, i.e. set of interfering pulse trains.

Table I: Five main sources of error in the clock offset due to the presence of motion. The estimated maximum errors are calculated for a closing velocity of +30 m/s and the current system parameters. The symbols are defined in the text.

| Source | Dependence | Maximum error (ps) |
|---|---|---|
| Fundamental Relativistic Breakdown in Reciprocity | $\frac{V}{c^2}L$ | 1 |
| Breakdown due to Asynchronous Sampling | $\frac{V}{c}\frac{1}{2\Delta f_r}$ | 20 |
| Delay-Doppler Coupling | $\left(\frac{V}{c}\right)2\pi\nu_c\left(\beta_2^{path}+\beta_2^{RF}+\frac{f_r}{\Delta f_r}\Delta\beta_2^{Combs}\right)$ | 1 |
| Velocity-Dependent Transceiver Calibration | $\frac{V}{c^2}L_{\text{transceiver}}$ | 0.003* |
| Periodic Waveform Ambiguities | $\frac{1}{f_r}$ | 5000** |

*$L_{\text{transceiver}}$ ~ *10 m*

**Assuming an error of one pulse

## III.B Detailed Derivation of Time-Frequency Comparison/Synchronization Equations

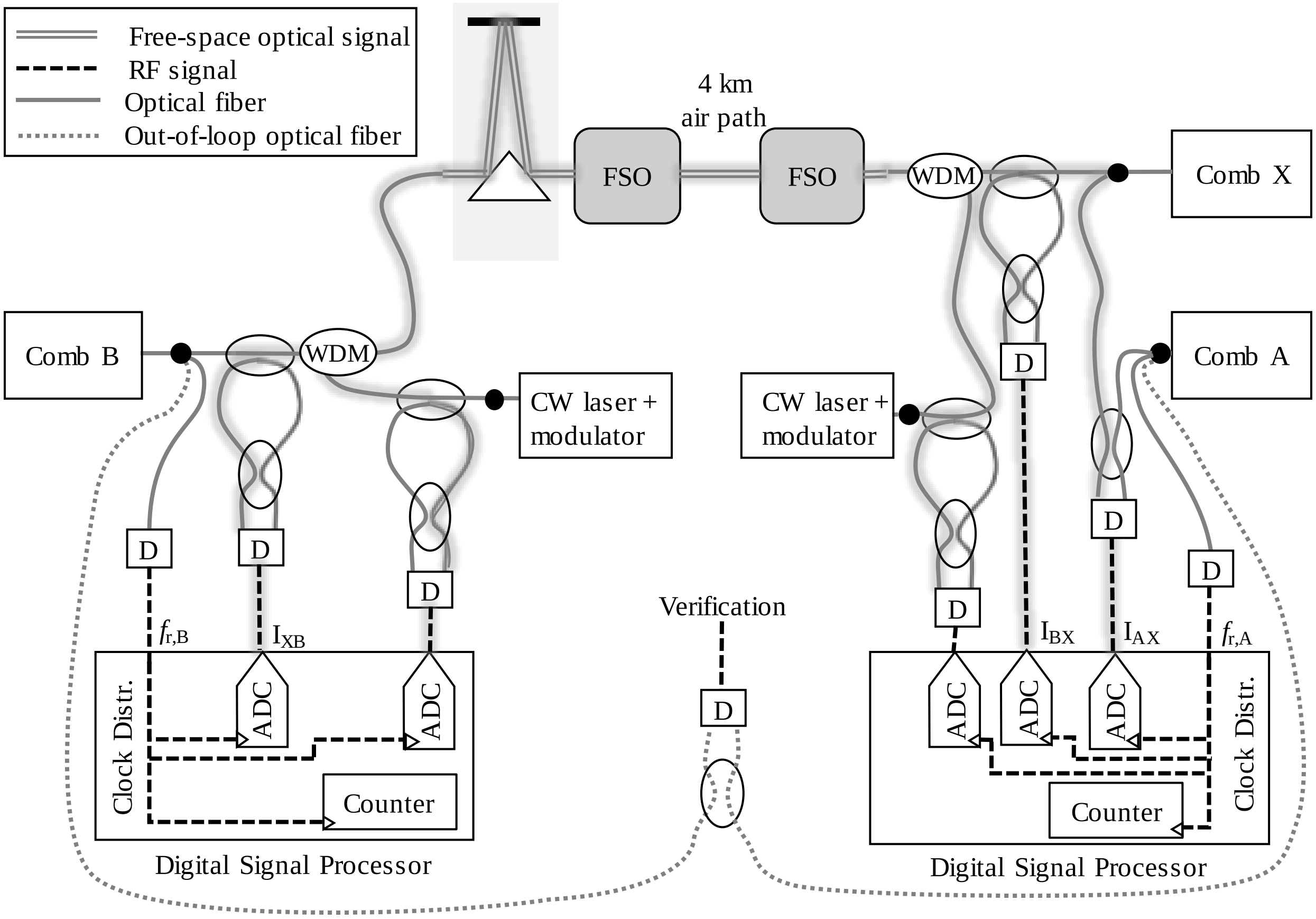

Figure 4: Detailed schematic of master and remote sites. WDM: wavelength division multiplexor, ADC: Analog-to-Digital Converter, white ovals: fiber couplers, black dots: injection sites for transceiver calibration via OTDR, grey shading: one example set of delays (see Section III.B.3).

We now give a detailed derivation specific to the comb-based O-TWTFT. An overall description of the system was already provided in the introduction to Section II. The accompanying Fig. 2 above presents a high-level schematic of the system, while Fig. 4 is a more detailed schematic that connects the quantities discussed in the derivation with the physical layout. As emphasized in Fig. 4, at each site, all signals are digitized and processed in a real-time signal processor whose clock is driven by the local comb A or B. Therefore, all digitized samples are recorded on the local timebase, albeit with additional timing jitter and timing delays inherent in transferring the optical pulse train timing to the ADC clock (that must be calibrated and removed.) Indeed, as shown in Fig. 4, there are multiple delays between various detected signals within the transceiver. For the derivation, we initially ignore these delays and treat all measurements as occurring at a single reference point. Later, we discuss the transceiver calibration that adjusts the timing of these signals to the reference plane, effectively applying the calibration term, $\Delta T_{\text{cal}} + \left(V/c\right)\Delta T_{\text{cal}}^{V}$ , discussed earlier.

III.B.1 *Local Timebase*

The master frequency comb A's field at Site A's reference plane, $z = z_{\text{A}}$, is

$$E_{\text{A}}\left(t, z_{\text{A}}\right) = e^{i2\pi\tilde{\nu}_{\text{A}}t}\sum\nolimits_{m} E_{\text{A},m} e^{im\Phi_{\text{A}}\left(t,z_{\text{A}}\right)} \tag{2}$$

where $t$ is a general oracle time (a purely notational/mathematical convenience), the integer $m$ labels the comb tooth number, $\tilde{\nu}_{\text{A}}$ is the frequency of some central comb tooth, $E_{\text{A},m}$ is the amplitude of the $m$th comb tooth, and $\Phi_{\text{A}}\left(t, z\right)$ is the phase of comb A's pulse train at a position $z$ and oracle time $t$. We write identical expressions for the remote and transfer combs with the subscript "A" replaced by "B" and "X", respectively. At site B, we define a reference plane $z = z_{\text{B}}$. The repetition rate of Combs A and B are $f_{r,\text{A}}\left(t\right) = \left(2\pi\right)^{-1} d\Phi_{\text{A}}\left(t, z_{\text{A}}\right)/dt$ and $f_{r,\text{B}}\left(t\right) = \left(2\pi\right)^{-1} d\Phi_{\text{B}}\left(t, z_{\text{B}}\right)/dt$, respectively, as measured against the oracle timebase. Against their own timebases, both repetition rates are exactly the nominal repetition frequency $\hat{f}_r$ by definition, in this case, Eq. (2) is the usual comb equation with exponents $2\pi m\hat{f}_r t + \Phi\left(0, z_A\right)$. Comb X's repetition rate $f_{r,\text{X}}\left(t\right) = \left(2\pi\right)^{-1} d\Phi_{\text{X}}\left(t, z\right)/dt$ is offset by $\sim\Delta f_r$, the nominal difference in repetition frequencies. This offset repetition frequency $\Delta f_r$ sets the fundamental update rate of the overall measurement and is 2 kHz here. Throughout, we assume $f_r\ (t)$ is close to $\hat{f}_r$ and varies slowly compared to $1/\Delta f_r$ . As shown in Figure 2a, the repetition rate of each clock comb A or B is phase locked to the underlying optical oscillator at its site.

In the above expression, $t$ is some inaccessible oracle time. The measurable time at each site is defined through the comb phase. For site A, the phase of comb A, $\Phi_{\text{A}}(t, z_{\text{A}})$ defines the timebase as indicated in the following equivalent expressions:

$$\begin{aligned}\Phi_{\text{A}}\left(t, z_{\text{A}}\right) &\equiv 2\pi \hat{f}_r t_{\text{A}}\left(t\right) \\ &= 2\pi \hat{f}_r \left\{t + \Delta t_{\text{A}}\left(t\right)\right\}. \\ &= 2\pi k_{\text{A}}\end{aligned} \tag{3}$$

The first expression of Eqn. (3) expresses the direct relationship between site A's timebase $t_{\text{A}}\left(t\right)$ and comb A's phase. In other words, a "tick" in the timebase occurs at every integer multiple (of $2\pi$) of the phase -- or equivalently at the arrival of an optical pulse at the reference plane – and we define the time interval between ticks as $1/\hat{f}_r$ according to that clock. The phase is a continuous function so that this timebase is well defined in the intervals between ticks. In general, we use the phase, $\Phi_{\text{A}}\left(t, z_{\text{A}}\right)$, rather than $t_{\text{A}}\left(t\right)$ to describe the timebase, which avoids notational and Doppler-related complexities. The second expression of Eqn. (3) relates the comb A timebase to oracle time through a slowly varying time-offset $\Delta t_{\text{A}}\left(t\right)$. (In the Methods of Ref. (), this quantity appears as $\tau_A$). Finally, the third expression of Eqn. (3) relates the comb A timebase to the sample number, $k_{\text{A}}$, of the analog-to-digital converter (ADC) at site A (which is clocked by the comb A pulse train as shown in Fig. 4). In reality, there is excess timing jitter and an additional time offset between comb A's phase and $k_{\text{A}}$, but we ignore these factors until the calibration discussion. To reiterate, integer values of ADC sample number, $k_{\text{A}}$, correspond to the $k_{\text{A}}$th pulse arrival at $z = z_{\text{A}}$ and to $\Phi_A\left(t, z_A\right) = 2\pi k_A$. As the phase is continuous, so too $k_A$ is not restricted to integer values and we will consider fractional values retrieved, for example, from fitting a peak with subsample precision. At Site B, all the above equations apply after replacing the subscript A with B.

Our goal is to calculate the time offset between sites, i.e. $\Delta t_{AB} \equiv \Delta t_A - \Delta t_B = \left(2\pi \hat{f}_r\right)^{-1}\left[\Phi_A\left(t, z_A\right) - \Phi_B\left(t, z_B\right)\right]$. The simplest approach would be direct two-way transfer of Combs A and B between the sites followed by direct detection, as implied in Section III.A. However, the timing signals then have picosecond-level jitter/systematics because of photodetection and the ADC sample clock jitter and systematics. We can avoid this uncertainty through heterodyne detection between the optical pulses themselves. However, the short pulses from combs A and B would rarely overlap (unless the time-of-flight was an exact multiple of $\hat{f}_r^{-1}$). Hence, we introduce the transfer comb X that runs at an offset repetition rate as illustrated in Fig. 2. We measure the phase difference between Comb X and Comb A at the master site, and between Comb X and Comb B both at the remote and master site. From these data, we extract the desired phase difference $\Phi_A\left(t, z_A\right) - \Phi_B\left(t, z_B\right)$.

III.B.2 *Computation of the Clock Offset*

The three heterodyne signals between the combs effectively measure their phase differences, as illustrated in Fig. 5. The pulsed nature of the comb signal provides a much more precise measurement of the phase crossings (i.e. points of equal phase modulo $2\pi$) between oscillators, but interpolation is required to evaluate the phase difference during the intervening period. Figure 5c shows the phase difference between combs A and X, evaluated at the master site. We similarly measure the phase difference between combs B and X at each site.

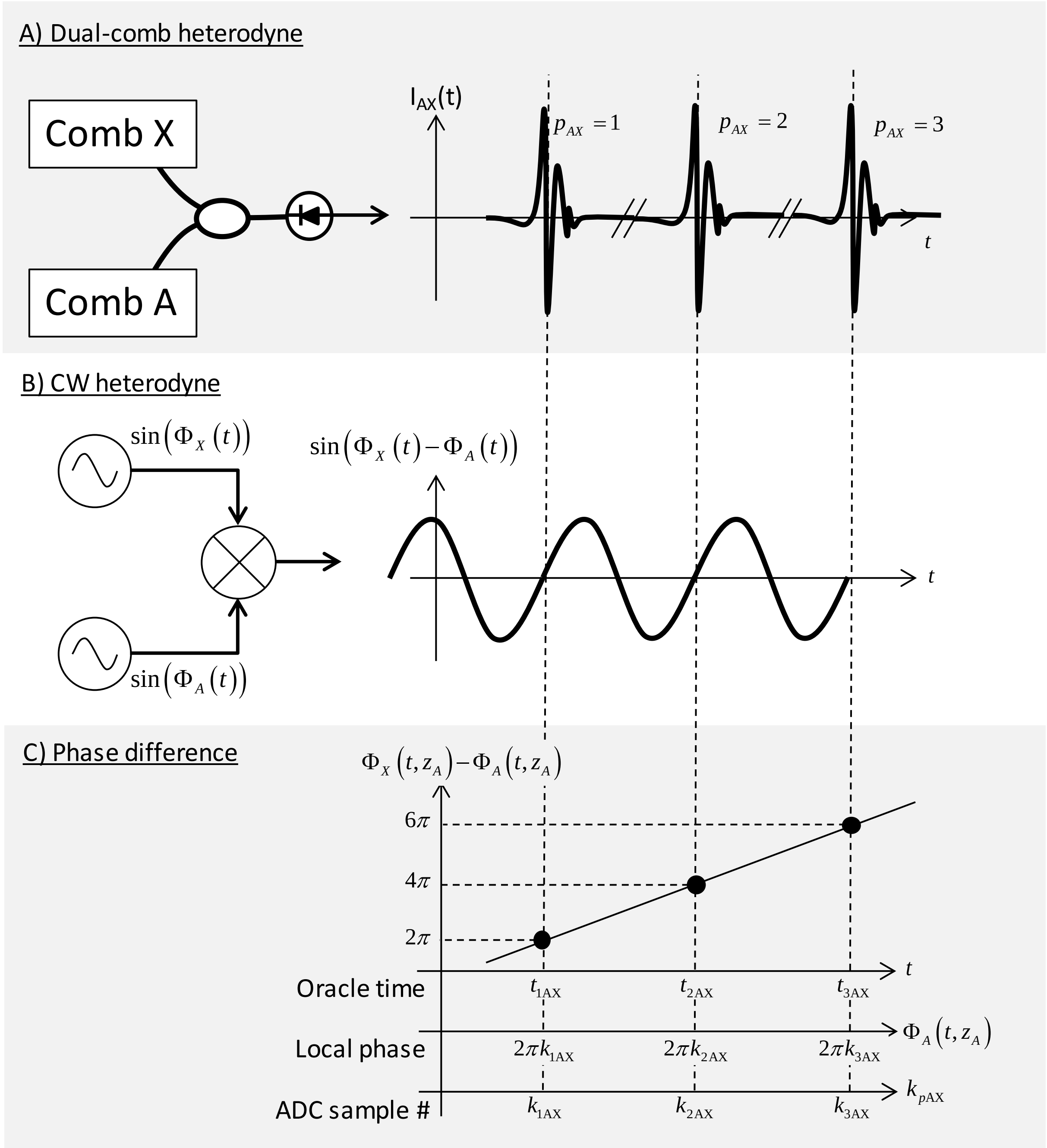

Figure 5: Diagram illustrating the relationship between the interferogram produced by (a) the interference of two combs and (b) the analogous heterodyne mixing of continuous-wave oscillators. (c) The peak of the interferograms between the combs yield their associated phase difference with orders-of-magnitude higher precision than if the detected repetition rates were instead measured as in (b). This higher precision is a consequence of the shot-noise limited signal-to-noise ratio and the ~ 1 THz measurement bandwidth set by the optical pulse bandwidth. As indicated in the bottom graph, we can view the phase offset, alternately, against "oracle time", which is mathematically convenient but experimentally inaccessible, or the local timebase, evaluated in terms of the local optical phase or the local ADC sample number. Note the $k_{AX}$ are not integers.

Mathematically, the intensity of the heterodyne signal between the master (A) and transfer (X) combs (i.e. the master-transfer interferogram) is

$$I_{\mathrm{AX}}\left(t,z_{\mathrm{A}}\right)=e^{i2\pi(\tilde{\nu}_{\mathrm{X}}-\tilde{\nu}_{\mathrm{A}})t}\sum_{m'}E^{*}_{\mathrm{A},m'}e^{-im'\Phi_{\mathrm{A}}\left(t,z_{\mathrm{A}}\right)}\sum_{m}E_{\mathrm{X},m}e^{im\Phi_{\mathrm{X}}\left(t,z_{\mathrm{A}}\right)}+c.c.. \tag{4}$$

(It is actually measured at the local ADC indicated in Fig. 4, but we consider it here at $z = z_{\mathrm{A}}$). We select similar nominal central frequencies so that low-pass filtering retains only terms for which $m=m'$. Figure 6 shows an example measured interferogram.

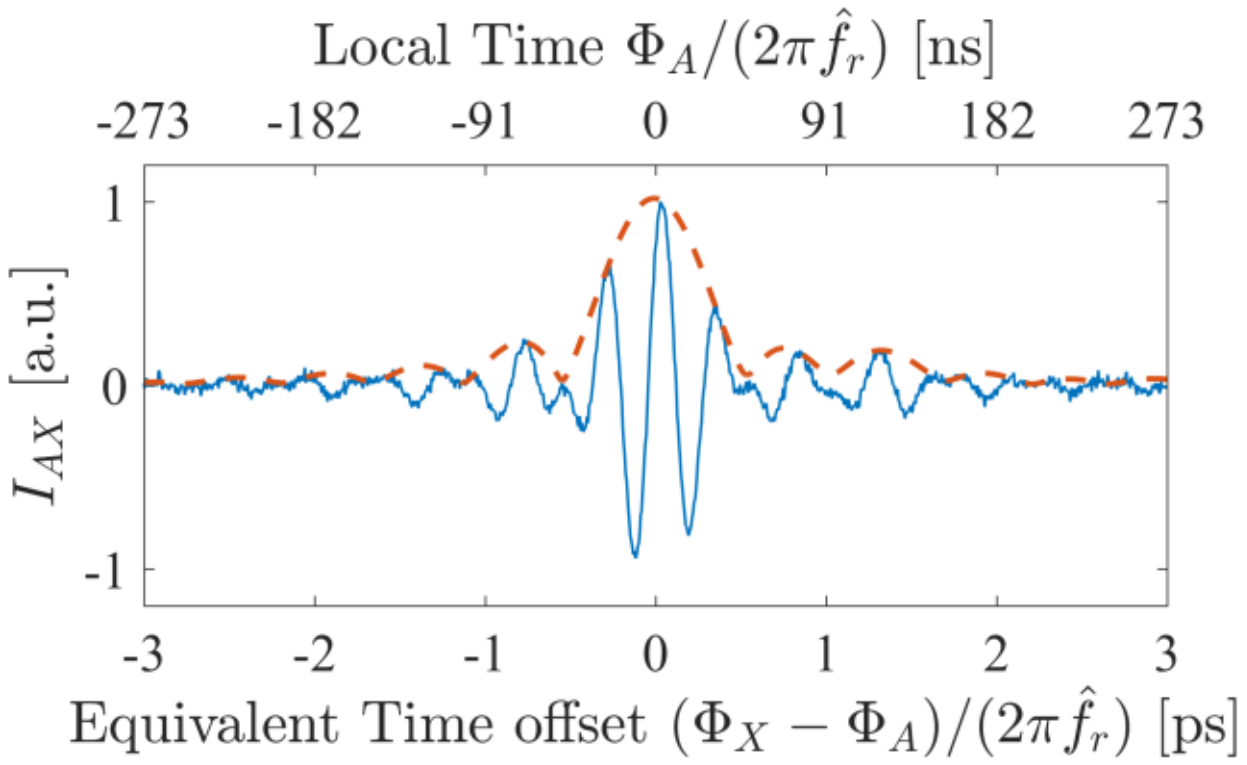

Figure 6: Example digitized interferogram between comb A and comb X (blue line). To find its center, it is first filtered by a matched filter, then a Hilbert transform is applied to generate an envelope function (red dashed line) followed by a subsample interpolation to find its precise peak position. The other interferograms require a more involved procedure (cross-ambiguity function search) to find their peaks' position due to Doppler shifts.

With appropriate substitution of the subscripts, we write similar expressions for the remote-transfer interferogram, $I_{\mathrm{BX}}(t,z_{\mathrm{A}})$ and the transfer-remote interferogram, $I_{\mathrm{XB}}(t,z_{\mathrm{B}})$. Dropping the carrier term and assuming a flat detector response (see Appendix A), the three interferograms are

$$\begin{aligned}
I_{\mathrm{AX}}\left(t,z_{\mathrm{A}}\right)&=\sum_{m}E_{\mathrm{X},m}E^{*}_{\mathrm{A},m}e^{im\left\{\Phi_{\mathrm{X}}\left(t,z_{\mathrm{A}}\right)-\Phi_{\mathrm{A}}\left(t,z_{\mathrm{A}}\right)\right\}}\\
I_{\mathrm{BX}}\left(t,z_{\mathrm{A}}\right)&=\sum_{m}E_{\mathrm{X},m}E^{*}_{\mathrm{B},m}e^{im\left\{\Phi_{\mathrm{X}}\left(t,z_{\mathrm{A}}\right)-\Phi_{\mathrm{B}}\left(t,z_{\mathrm{A}}\right)\right\}}\\
I_{\mathrm{XB}}\left(t,z_{\mathrm{B}}\right)&=\sum_{m}E^{*}_{\mathrm{B},m}E_{\mathrm{X},m}e^{im\left\{\Phi_{\mathrm{X}}\left(t,z_{\mathrm{B}}\right)-\Phi_{\mathrm{B}}\left(t,z_{\mathrm{B}}\right)\right\}}
\end{aligned} \tag{5}$$

These sums describe a series of interferograms, or peaks, at times whenever the phase difference is zero modulo $2\pi$, as illustrated in Fig. 5. They repeat at the difference in repetition rates, $\Delta f_r$. Consider the stream of interferograms represented by $I_{\mathrm{AX}}$. Let us introduce the integer $p_{\mathrm{AX}}$ that counts successive interferogram peaks that occur at oracle times $t = t_{p\mathrm{AX}}$ so that we have $\Phi_{\mathrm{X}}\left(t_{p\mathrm{AX}}, z_{\mathrm{A}}\right) - \Phi_A\left(t_{p\mathrm{AX}}, z_{\mathrm{A}}\right) = 2\pi p_{\mathrm{AX}}$. Of course, we do not have access to these oracle times, $t_{p\mathrm{AX}}$. Rather, we have access to the interferogram peak locations with respect to the local ADC clock: $\Phi_{\mathrm{A}}\left(t_{p\mathrm{AX}}, z_{\mathrm{A}}\right) \equiv 2\pi k_{p\mathrm{AX}}$. (See Figure 5c.) We record the ADC sample numbers $k_{p\mathrm{AX}}$ corresponding to the $p_{\mathrm{AX}}$ interferogram with sub-sample precision by applying a matched filter to the incoming interferogram signal, calculating the Hilbert transform to find the envelope, and fitting the peak. (See Fig. 6.) The end result is set of paired values $\left\{p_{AX}, k_{pAX}\right\}$.

For the interferograms $I_{\mathrm{BX}}$ and $I_{\mathrm{XB}}$, we introduce the analogous integers $p_{\mathrm{BX}}$ and $p_{\mathrm{XB}}$ that count successive interferogram peaks occurring at oracle times $t_{p\mathrm{BX}}$ and $t_{p\mathrm{XB}}$. For $I_{\mathrm{BX}}$, we find the interferogram peak location as $k_{p\mathrm{BX}}$ against site A's ADC clock, whereas for $I_{\mathrm{XB}}$, we find the interferogram peak location as $k_{p\mathrm{XB}}$ against site B's ADC clock. In these cases, the incoming comb light and thus interferograms suffer from Doppler shifts. Therefore, we cannot find the peak locations by the same matched-filter approach as for the $k_{p\mathrm{AX}}$ as this would incur intolerable systematics due to coupling between the extracted peak values and the Doppler shifts due to dispersion. (See Section III.A.3.) Instead, we use a cross-ambiguity function search, as described in Appendix A. The end results is again two sets of paired values, $\left\{p_{BX}, k_{pBX}\right\}$ and $\left\{p_{XB}, k_{pXB}\right\}$.

We connect the recorded pairs of data from the interferogram peaks with the comb phases as follows.

On site A, from the master-transfer (AX) interferogram,

$$\Phi_{\mathrm{X}}\left(t_{p\mathrm{AX}}, z_{\mathrm{A}}\right)-\Phi_{\mathrm{A}}\left(t_{p\mathrm{AX}}, z_{\mathrm{A}}\right)=2\pi p_{\mathrm{AX}} , \tag{6}$$

$$\Phi_{\mathrm{A}}\left(t_{p\mathrm{AX}}, z_{\mathrm{A}}\right)\equiv 2\pi k_{p\mathrm{AX}} . \tag{7}$$

On site A, from the remote-transfer (BX) interferogram,

$$\Phi_{\mathrm{X}}\left(t_{p\mathrm{BX}}, z_{\mathrm{A}}\right)-\Phi_{\mathrm{B}}\left(t_{p\mathrm{BX}}, z_{\mathrm{A}}\right)=2\pi p_{\mathrm{BX}} , \tag{8}$$

$$\Phi_{\mathrm{A}}\left(t_{p\mathrm{BX}}, z_{\mathrm{A}}\right)\equiv 2\pi k_{p\mathrm{BX}} . \tag{9}$$

On site B, from the transfer-remote (XB) interferogram,

$$\Phi_{\mathrm{X}}\left(t_{p\mathrm{XB}}, z_{\mathrm{B}}\right)-\Phi_{\mathrm{B}}\left(t_{p\mathrm{XB}}, z_{\mathrm{B}}\right)=2\pi p_{\mathrm{XB}} . \tag{10}$$

$$\Phi_{\mathrm{B}}\left(t_{p\mathrm{XB}}, z_{\mathrm{B}}\right)\equiv 2\pi k_{p\mathrm{XB}} . \tag{11}$$

In writing Eqns. (6) through (11), the comb phases appear on the left hand side and measured quantities appear on the right hand side. Again, these equations for now ignore any offsets due to propagation delays internal to each site and timing noise. (Section III.B.3 discusses the system calibration.)

The integers $p_{\mathrm{AX}}$, $p_{\mathrm{BX}}$, and $p_{\mathrm{XB}}$ represent the number of excess pulses that the transfer comb has accumulated compared to the master or remote comb since the start of each site's pulse counter. $p_{\mathrm{AX}}$ is straightforwardly counted as the co-located master and transfer combs yields a stable, reliably measurable $I_{\mathrm{AX}}$ that peaks for each and every excess pulse from the transfer comb. However, turbulence-induced fades can lead to missing interferograms for $I_{\mathrm{BX}}$ and $I_{\mathrm{XB}}$. Therefore, $p_{\mathrm{BX}}$ and $p_{\mathrm{XB}}$ must be resolved using the communications-channel-based TWTFT system as described in Appendix B.

We now rewrite these equations to (i) effectively eliminate the transfer comb phase and (ii) formally obtain equations as in two-way time transfer. This elimination of the transfer comb phase is achieved by using Eqn. (6) and a linear expansion of the phase to map $\Phi_X$ to $\Phi_A$ as

$$\Phi_X\left(t, z_A\right) = \Phi_A\left(t, z_A\right) + 2\pi p_{AX} + 2\pi\Delta f_r \left\{t - t_{pAX}\right\} \tag{12}$$

where $t$ is near $t_{pAX}$. In the implementation, this mapping is done through interpolation and does not require knowledge of $\Delta f_r$. Note the right-hand side includes oracle times but only as a difference, and thus the offset between oracle time and the local site A timebase, $\Delta t_A$, drops out. We will use Eq. (12) liberally below.

To formally obtain the two-way time-transfer equations, we identify two events that occur at every update time interval $1/\Delta f_r$ (assuming no turbulence fades):

Event 1: Effective transmission of the comb A time (or phase) to site B, as recorded by the $p_{XB}$th peak of the $I_{XB}$ interferogram at oracle time $t = t_{pXB}$. In the conventional two-way time transfer, we would record the departure time $T_{AA}$ from site A (as measured against site A's timebase) and the arrival time $T_{AB}$ at site B (as measured against site B's timebase). In analogy, for Event 1, we formally define:

$$\begin{aligned} T_{AB} &\equiv \left(2\pi \hat{f}_r\right)^{-1} \Phi_B\left(t_{pXB}, z_B\right) \\ T_{AA} &\equiv \left(2\pi \hat{f}_r\right)^{-1} \Phi_A\left(t_{pXB} - T_{A\to B}\left(t_{pXB}\right), z_A\right) \end{aligned} \tag{13}$$

where we convert from phase to time through the nominal repetition rate, $\hat{f}_r$ and $T_{A\to B}\left(t_{pXB}\right)$ is the time-of-flight for a signal that arrives at site B at oracle time $t_{pXB}$. Therefore, we have a pair of $T_{AB}$ and $T_{AA}$ associated with each $p_{XB}$.

Event 2 Effective transmission of the comb B time (or phase) to site A, as recorded by the $p_{BX}$th peak of the $I_{BX}$ interferogram at oracle time $t = t_{pBX}$. Again, in the conventional two-way time transfer, we would record the departure time $T_{BB}$ from site B (as measured against site B's timebase) and the arrival time $T_{BA}$ at site A (as measured against site A's timebase). In analogy, for Event 2 we formally define:

$$
\begin{aligned}
\mathrm{T}_{\mathrm{BA}} &\equiv \left(2\pi \hat{f}_r\right)^{-1} \Phi_A\left(t_{p\mathrm{BX}}, z_{\mathrm{A}}\right) \\
\mathrm{T}_{\mathrm{BB}} &\equiv \left(2\pi \hat{f}_r\right)^{-1} \Phi_B\left(t_{p\mathrm{BX}} - T_{\mathrm{B}\to\mathrm{A}}\left(t_{p\mathrm{BX}}\right), z_{\mathrm{B}}\right)
\end{aligned}
\qquad (14)
$$

where $T_{\mathrm{B}\to\mathrm{A}}\left(t_{p\mathrm{BX}}\right)$ is the time of flight for an event that arrives at site A at oracle time $t_{p\mathrm{BX}}$. We have a pair of $T_{BA}$ and $T_{BB}$ associated with each $p_{BX}$.

We can connect these definitions Eqns. (13) and (14) with the actual measurements, Eqns. (6) through (11) as (see Appendix C for derivation):

$$
\begin{aligned}
&\hat{f}_r \mathrm{T}_{\mathrm{AA}} = k_{p\mathrm{XB}} - \tfrac{\Delta f_r}{\hat{f}_r + \Delta f_r}\left(k_{p\mathrm{XB}} - k_{p\mathrm{AX}} + p_{\mathrm{XB}} - p_{\mathrm{AX}}\right) + p_{\mathrm{XB}} - p_{\mathrm{AX}} && \text{a)} \\
&\hat{f}_r \mathrm{T}_{\mathrm{AB}} = k_{p\mathrm{XB}} && \text{b)} \\
&\hat{f}_r \mathrm{T}_{\mathrm{BB}} = k_{p\mathrm{BX}} + \tfrac{\Delta f_r}{\hat{f}_r}\left(k_{p\mathrm{BX}} - k_{p\mathrm{AX}}\right) + p_{\mathrm{AX}} - p_{\mathrm{BX}} && \text{c)} \\
&\hat{f}_r \mathrm{T}_{\mathrm{BA}} = k_{p\mathrm{BX}} && \text{d)}
\end{aligned}
\qquad (15)
$$

We can now find the time difference by use of the standard two-way time-transfer equation, e.g. Eqn. (1),

$$
\Delta t_{\mathrm{AB}} = \frac{1}{2}\left[T_{\mathrm{AA}} - T_{\mathrm{AB}} - T_{\mathrm{BB}} + T_{\mathrm{BA}}\right] + \frac{1}{2}\left[T_{\mathrm{A}\to\mathrm{B}}\left(t_{p\mathrm{XB}}\right) - T_{\mathrm{B}\to\mathrm{A}}\left(t_{p\mathrm{BX}}\right)\right] + \Delta T_{\mathrm{cal}} + \left(V/c\right)\Delta T_{\mathrm{cal}}^{V} \qquad (16)
$$

evaluated at the mean oracle time $t = \left(t_{p\mathrm{BX}} + t_{p\mathrm{XB}}\right)/2$ and where we have introduced a static, $\Delta T_{\mathrm{cal}}$, and velocity dependent, $\Delta T_{\mathrm{cal}}^{V}$, calibration term whose determination is discussed in Section III.B3.

This formula can be derived by use of Eqn. (3) and their analog at Site B, Eqn. (13), Eqn. (14), and the expansion $\Phi_{\mathrm{B}}\left(t_{p\mathrm{BX}}-T_{\mathrm{B}\to\mathrm{A}}\left(t_{p\mathrm{BX}}\right),z_{\mathrm{B}}\right)\approx\Phi_{\mathrm{B}}\left(t_{p\mathrm{BX}},z_{\mathrm{B}}\right)-2\pi\hat{f}_{r}T_{\mathrm{B}\to\mathrm{A}}\left(t_{p\mathrm{BX}}\right)$.

To evaluate this quantity, e.g. Eqn. (16), we require the middle term, which is the non-reciprocal time-of-flight. Based on the discussion in Section III.A.1 and III.A.2 as well as Refs. [26,27], for our geometry

$$T_{\mathrm{A}\to\mathrm{B}}\left(t_{p\mathrm{XB}}\right)-T_{\mathrm{B}\to\mathrm{A}}\left(t_{p\mathrm{BX}}\right)=\frac{V}{c}\Delta t_{\mathrm{async}}+\frac{V}{c^{2}}\left[L_{\mathrm{A}}-L_{\mathrm{B}}\right] \tag{17}$$

which assumes a mirror moving at closing velocity $V/2$ located at a distance $L_{\mathrm{A}}(t)$ from site A and $L_{\mathrm{B}}(t)$ from site B. Note that in this geometry, the difference $L_{\mathrm{A}}-L_{\mathrm{B}}$ is time independent and must be roughly calibrated by measuring the distances for a shorted link. To generalize to the alternate scenario of a stationary clock A and a moving clock B, we would replace Eqn. (17) with $T_{\mathrm{A}\to\mathrm{B}}\left(t_{p\mathrm{XB}}\right)-T_{\mathrm{B}\to\mathrm{A}}\left(t_{p\mathrm{BX}}\right)=\frac{V}{c}\left[\Delta t_{\mathrm{async}}+T_{\mathrm{B}\to\mathrm{A}}\left(t_{p\mathrm{BX}}\right)\right]$ and in addition introduce a time dilation term. Note that the asynchronous sampling time offset, $\Delta t_{\mathrm{async}}=T_{\mathrm{AB}}-T_{\mathrm{BA}}+\Delta t_{\mathrm{AB}}$, defined initially in Section III.A.2 notably depends on the clock offset so that a solution of (17) requires solving for $\Delta t_{\mathrm{AB}}$.

However, we first require the instantaneous closing velocity, $V$, which is assumed slowly varying on the timescale of $1/\Delta f_{r}$. It is computed from the derivative of the timestamps and is given by

$$\frac{V}{c}=1-\sqrt{\frac{\dot{T}_{\mathrm{BB}}}{\dot{T}_{\mathrm{BA}}}\frac{\dot{T}_{\mathrm{AA}}}{\dot{T}_{\mathrm{AB}}}}, \tag{18}$$

where each derivative is computed using a 3-point numerical derivative centered on the right time, which is accurate up to second order. This equation follows from Eqns. (13) and (14) and simply reflects the Doppler shift on the received pulse trains, e.g. $\dot{T}_{\mathrm{BB}} = \left(2\pi \hat{f}_r\right)^{-1} d\Phi_{\mathrm{B}}\left(t - T_{\mathrm{B}\to\mathrm{A}}\left(t\right), z_{\mathrm{B}}\right) \Big/ dt = \left[1 - V/c\right] f_{r,\mathrm{B}}\left(t\right) \Big/ \hat{f}_r$ .

Solving Eqn. (16) by use of Eqns. (17) and the expression for $\Delta t_{\mathrm{async}}$ gives,

$$\Delta t_{\mathrm{AB}} = \frac{1}{2 - V/c} \left\{ T_{\mathrm{AA}} - T_{\mathrm{AB}} - T_{\mathrm{BB}} + T_{\mathrm{BA}} + 2\Delta T_{\mathrm{cal}} + \frac{V}{c}\left( \left[T_{\mathrm{AB}} - T_{\mathrm{BA}}\right] + \frac{1}{c}\left[L_{\mathrm{A}} - L_{\mathrm{B}}\right] + 2\Delta T_{\mathrm{cal}}^{V} \right) \right\} \quad (19)$$

or using Eqn. (15)a-d to express the clock offset directly in terms of the measured quantities,

$$\begin{aligned} \Delta t_{\mathrm{AB}} = \frac{1}{2 - V/c} & \left\{ \frac{\Delta f_r}{\hat{f}_r^{\,2}} \left[ k_{p\mathrm{AX}} - k_{p\mathrm{BX}} + \left(1 + \Delta f_r / \hat{f}_r\right)^{-1} \left( k_{p\mathrm{AX}} - k_{p\mathrm{XB}} - p_{\mathrm{XB}} + p_{\mathrm{AX}} \right) \right] \right. \\ & + \frac{1}{\hat{f}_r} \left[ p_{\mathrm{XB}} + p_{\mathrm{BX}} - 2 p_{\mathrm{AX}} \right] + 2\Delta T_{\mathrm{cal}} \\ & \left. + \frac{1}{\hat{f}_r} \frac{V}{c} \left( k_{p\mathrm{XB}} - k_{p\mathrm{BX}} + \hat{f}_r\, c^{-1} \left[ L_{\mathrm{A}} - L_{\mathrm{B}} \right] + 2 \hat{f}_r\, \Delta T_{\mathrm{cal}}^{V} \right) \right\} \end{aligned} \quad (20)$$

Alternatively, to connect with [16], we can retain an explicit time-of-flight term using Eqn. (33) from Appendix C and note that $t_{p\mathrm{AX}} - t_{p\mathrm{XB}} = f_r^{-1} k_{p\mathrm{AX}} - f_r^{-1} k_{p\mathrm{XB}} + \Delta t_{\mathrm{AB}}$ to write

$$\begin{aligned} \Delta t_{\mathrm{AB}} = \frac{1}{2 - V/c + \Delta f_r / \hat{f}_r} & \left\{ \frac{\Delta f_r}{\hat{f}_r} \left\{ 2 \hat{f}_r^{-1} k_{p\mathrm{AX}} - \hat{f}_r^{-1} k_{p\mathrm{XB}} - \hat{f}_r^{-1} k_{p\mathrm{BX}} \right\} + \frac{\Delta f_r}{\hat{f}_r} T_{\mathrm{A}\to\mathrm{B}}\left(t_{p\mathrm{XB}}\right) \right. \\ & + \hat{f}_r^{-1} \left[ p_{\mathrm{XB}} + p_{\mathrm{BX}} - 2 p_{\mathrm{AX}} \right] + 2\Delta T_{\mathrm{cal}} \\ & \left. + \frac{V}{c} \left( \hat{f}_r^{-1} k_{p\mathrm{XB}} - \hat{f}_r^{-1} k_{p\mathrm{BX}} + c^{-1} \left[ L_{\mathrm{A}} - L_{\mathrm{B}} \right] + 2\Delta T_{\mathrm{cal}}^{V} \right) \right\} \end{aligned} . \quad (21)$$

We note that solving for $T_{\mathrm{A}\to\mathrm{B}}(t_{p\mathrm{XB}})$ to first order in velocity through another combination of our four effective timestamps yields

$$T_{\mathrm{A}\to\mathrm{B}}(t_{p\mathrm{XB}}) = \tfrac{1}{2}\left[T_{\mathrm{AB}} - T_{\mathrm{AA}} + T_{\mathrm{BA}} - T_{\mathrm{BB}}\right] + \tfrac{1}{4}\left(V/c\right)\left[T_{\mathrm{AB}} + T_{\mathrm{AA}} - T_{\mathrm{BA}} - T_{\mathrm{BB}}\right] + \tfrac{1}{2}\left(V/c^2\right)\left[L_{\mathrm{A}} - L_{\mathrm{B}}\right]$$ which in turn could be expressed in terms of measured quantities via Eqn. (15).

III.B.3 *Transceiver calibration*

As discussed previously, the above equations assume measurements at one site are made at exactly the same physical point. However, as shown in Fig. 4, each site is far from a single physical point. To obtain the delays, we implemented a custom Optical Time Domain Reflectometer (OTDR) with the FPGA-DSP platform. A cw laser was amplitude-modulated by a 5 ns pulse and injected at all possible fiber inputs to the system, indicated by the black dots of Fig. 4, while the digitized detected signal was measured at all possible locations. One such set of paths for a single injection point is indicated by the grey shading in Fig. 4. We also include an additional detector to record the launch time of the pulse. Each pair of launch and detection times yields one particular linear combination of the system delays, which is captured through a measurement matrix M with each row representing one measurement and each column representing one delay.

This system of equations alone was not sufficient to solve independently for all the possible delays (there are far too few measurements). Consequently, we built a second matrix C (Calibration) representing only the subset of linear combinations of delays required to adjust the various timestamps to relate them all to a single calibration point. The problem was thus reduced to finding a linear combination of our measurements that yielded each required calibration value, itself a linear combination of the physical delays of the system, which can be succinctly represented as solving the system: C = A*M, where each row of A contains the coefficients of the linear combination of measurements required to compute one calibration value. By using the same RF paths, clock distribution and ADCs as the synchronization measurements, the delays associated with the FPGA-DSP platform are taken into account. This step needs to only be performed once as long as the transceiver configuration is not altered. We could combine the various delays to generate the values for $\Delta T_{\text{cal}}$ and $\Delta T_{\text{cal}}^{V}$. However, in the actual implementation we simply apply separate correction terms to each of the measured values before combining them into Eqn. (19).

Note that these delays can change with time. In particular, environmentally-induced phase noise due to temperature fluctuations or vibrations will cause variations in the delay values and can appear as an error in clock offset. Fortunately, the impact of environmentally-induced noise on the RF cables and components is suppressed by $\Delta f_r / f_r$. The main concern is, therefore, temperature-induced variations in the various "non-reciprocal" fiber optic paths, e.g. a portion of the fiber inside the optical transceivers. (Note that fiber paths connecting the transceivers to the free-space optical telescopes are part of the link and thus any variations in the delay from these is reciprocal and thus removed.) These non-reciprocal fiber paths are thus minimized in length and housed in temperature-controlled aluminum boxes.

In the implementation of the calibration procedure, we do not include the integer number of pulse repetition periods in the delays between the comb signal and the ADC clock input, which ultimately clocks the FPGA sample counter. This delay only affects the value of the sample counter, which takes on an arbitrary value at the system boot time. They have no effect for a static situation. With motion, the sub-sample parts of these delays do matter. However, as the maximum potential residual uncertainty in the computed clock offset is $(V/c)*5\text{ ns} = 0.4\text{ fs}$ for our maximum speed of 24 m/s, we chose to leave them out of the initial calibration. As shown in Section V.B, the system operated below this maximum bias value and thus we could not see this contribution. For systems operating at speeds greater than 60 m/s, these delays must be coarsely calibrated to avoid the residual uncertainty exceeding 1 fs.

As discussed in Appendix A, to avoid Doppler-systematics, we must also compensate for the electrical dispersion in the detection chain. To estimate the required compensation filters, the system's electrical impulse response was measured by injecting pulses from an external frequency comb at a highly offset repetition rate to create occasional impulses (single-point interferograms). This optical injection technique of ultra-short pulses required no modification of the RF path, ensuring that all synchronization measurements used the identical RF signal path as the one seen by the interference signal.

In the case of our "Doppler simulator" rail, an additional calibration step must be performed to compute the initial distance from the remote and master clocks to the retroreflector, i.e. $c^{-1}\left[L_{\mathrm{A}}-L_{\mathrm{B}}\right]=c^{-1}\left[L_{\mathrm{A}}\left(0\right)-L_{\mathrm{B}}\left(0\right)\right]$. This distance needs to be known to 30 cm to achieve 100 as timing and was determined in the same OTDR measurement as described above.

# IV. Experimental implementation and testbed

Sections II provides a high-level view of the system with more details of the optical system given in Refs. [2–4,28]. Here, we highlight the modifications required to suppress the motion-related effects by the necessary three-four orders of magnitude. First, we physically altered the optical transceivers of Refs. [2–4,28] to reduce the dispersion and therefore the delay-Doppler coupling. In addition, as noted above, there was significant ancillary calibration hardware and firmware developed for the calibration. Second, while nearly invisible in the schematics of Figs. 2 and 4, the digital signal processing required a new architecture. We now use a much more flexible combination of an FPGA and a DSP to implement the equations derived in the previous section in real time at a 2-kHz update rate, thereby enabling real-time synchronization between remote sites with a 10-100 Hz effective bandwidth. Third, we discuss operation of the free-space link with either a rapidly swept rail-mounted retroreflector or with a quadcopter-mounted retroreflector.

## IV.A Optical Transceivers

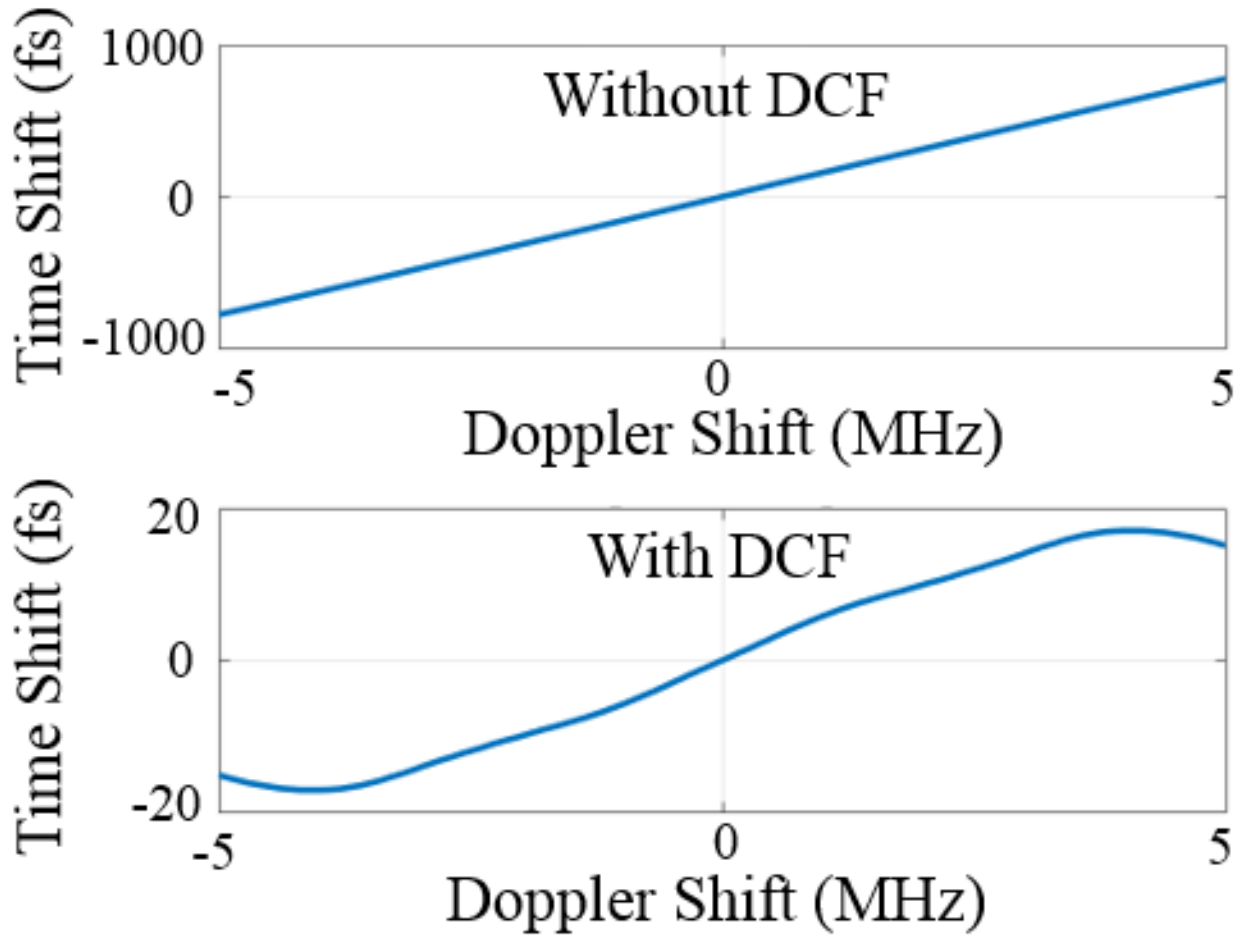

Figure 7: Measured time shift in the interferogram position as a function of Doppler shift, both without dispersion compensating fiber (DCF) and with DCF. The improvement with the addition of DCF is evident in the factor of 50 reduction in y-axis between the top and bottom plots. It is evident in the latter computation that a simple linear compensation for the delay-Doppler coupling is insufficient to maintain fs-level timing.

The optical transceivers at each site are very similar to that of Refs. [2,28]. However, unlike the case of a slowly-varying link, the total optical dispersion must be minimized to mitigate the delay-Doppler coupling as described in Section III.A.3. Specifically, the local comb pulse train needs to experience close to the same dispersion as the incoming pulse train in order to avoid the largest delay-Doppler systematic of magnitude, $\left( f_r / \Delta f_r \right) 2\pi \nu_c \Delta\beta_2^{Combs} V/c$. To this end, we add polarization-maintaining dispersion compensating fiber (DCF) at the output of each site's transceiver. This fiber is in the common path of both combs B and X. It compensates for the dispersion accumulated in the fiber optic paths from the transceivers to the free-space optical terminals and to the Doppler simulator. The home-built DCF module had a 4-dB insertion loss, but the low insertion loss (1.5 dB) of the free-space optical terminals [20] still enabled operation at 4 km. (Lower loss DCF modules are also commercially available.) Figure 7shows the delay-Doppler coupling in terms of systematic time shift of the extracted interferogram peak location versus Doppler shift, both before and after the insertion of the DCF module. The DCF module reduced the delay-Doppler coupling by a factor of 30. As the effects of second-order dispersion, $\Delta\beta_2^{Combs}$, is minimized, it is clear from Fig. 7 that higher order dispersion effects must be considered. To counteract systematics from all dispersion effects and to avoid further calibrations associated with this effect, we perform a two-dimensional search of the cross-ambiguity function. This approach is discussed in Appendix A. Figure 8 compares the cross-ambiguity function before and after the insertion of the DCF.

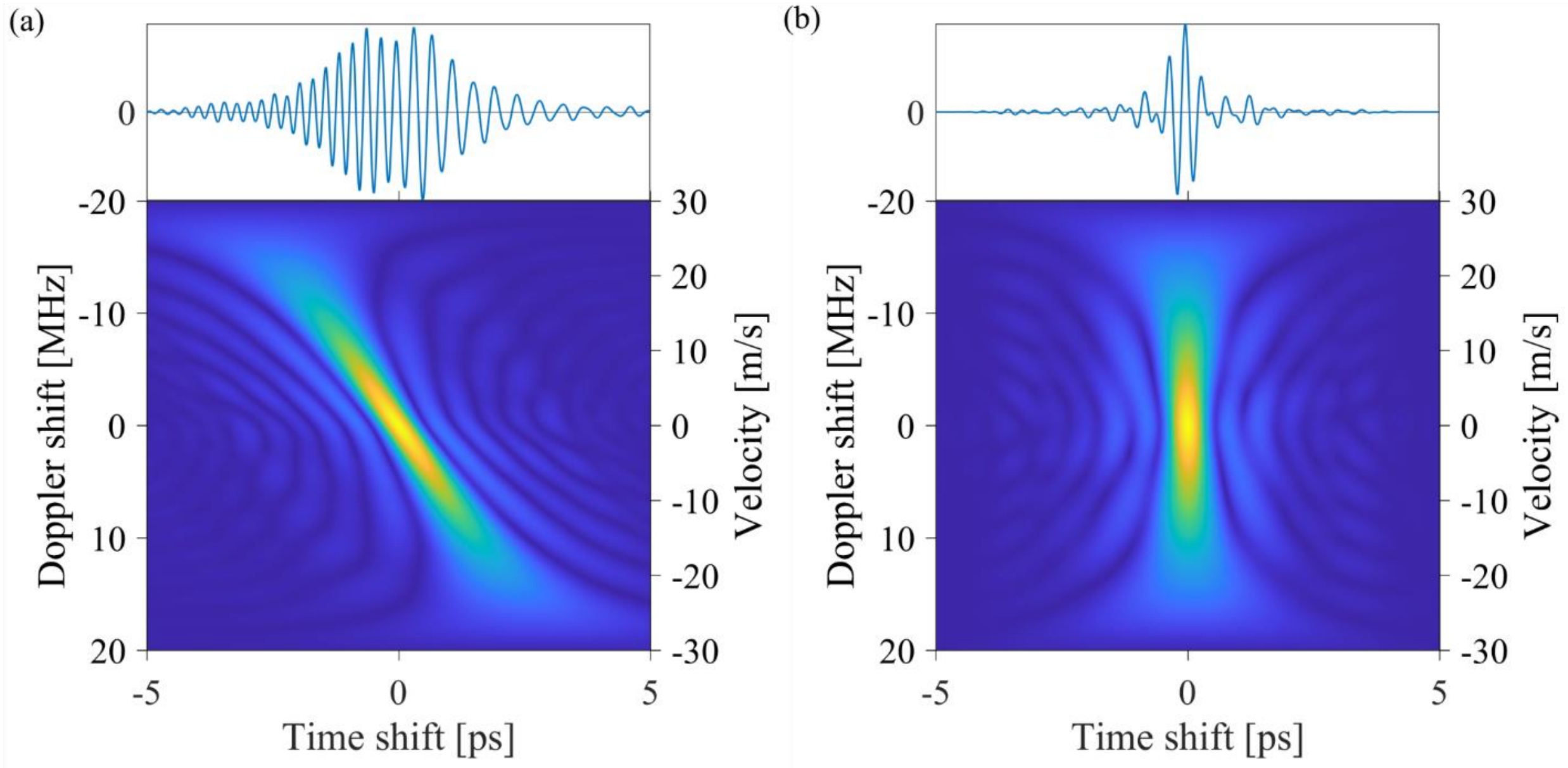

Figure 8: Delay-Doppler coupling. The specific waveform dispersion causes coupling between the closing velocity (Doppler shift) and the measured arrival time of the interferogram as recorded before and after the insertion of the DCF. (a) High differential dispersion, $\Delta\beta_2^{Combs}$, leads to a chirped interferogram (top figure) and a large delay-Doppler coupling as illustrated by the cross-ambiguity function (bottom figure). The amplitude of the cross-ambiguity function is shown on an arbitrary linear scale with warmer colors indicating higher intensity (greater correlation). (b) Low differential dispersion achieved through the insertion of dispersion compensating fiber reduces the interferogram chirp (top figure) and the delay-Doppler coupling (bottom figure). A vertical line would reflect zero delay-Doppler coupling.

## IV.B Digital Signal Processing

The real-time digital signal processing enables all processing steps, from initial interferogram detection to clock offset computation, to occur within a single interferogram period, $1/\Delta f_r \sim 500\ \mu s$. The computed clock offset is then fed to a Kalman-filter-based loop filter to allow for feedback to comb B with a synchronization bandwidth of 10 – 100 Hz. Here, we provide more details on the implementation of the digital signal processing with a high-level view of the signal processing on site B captured in Figure 9. The digital signal processor (DSP) is a 1.25 GHz multicore DSP with hardware floating-point support, running a bare-metal application. It interfaces with the Field Programmable Gate Array (FPGA), which is a Virtex 7 XC7VX485T [29] and contains 486k logic cells and 2.8k dedicated hardware multipliers running at 200-MHz clock rate.

From the comb-based system, we detect three heterodyne signals corresponding to the three series of interferograms, two on the master site and one on the remote site. All three heterodyne signals are low-pass filtered before digitization by the ADC at a 200 MHz sampling rate. The time series of the heterodyne signals then passes through a wideband digital filter, which compensates for the electrical dispersion (see Section III.B.3), and then a Hilbert-transform filter. When the amplitude of the Hilbert-filtered signal is above a given threshold, the processor passes a short data window (512 samples) around the peak to the DSP for further processing.

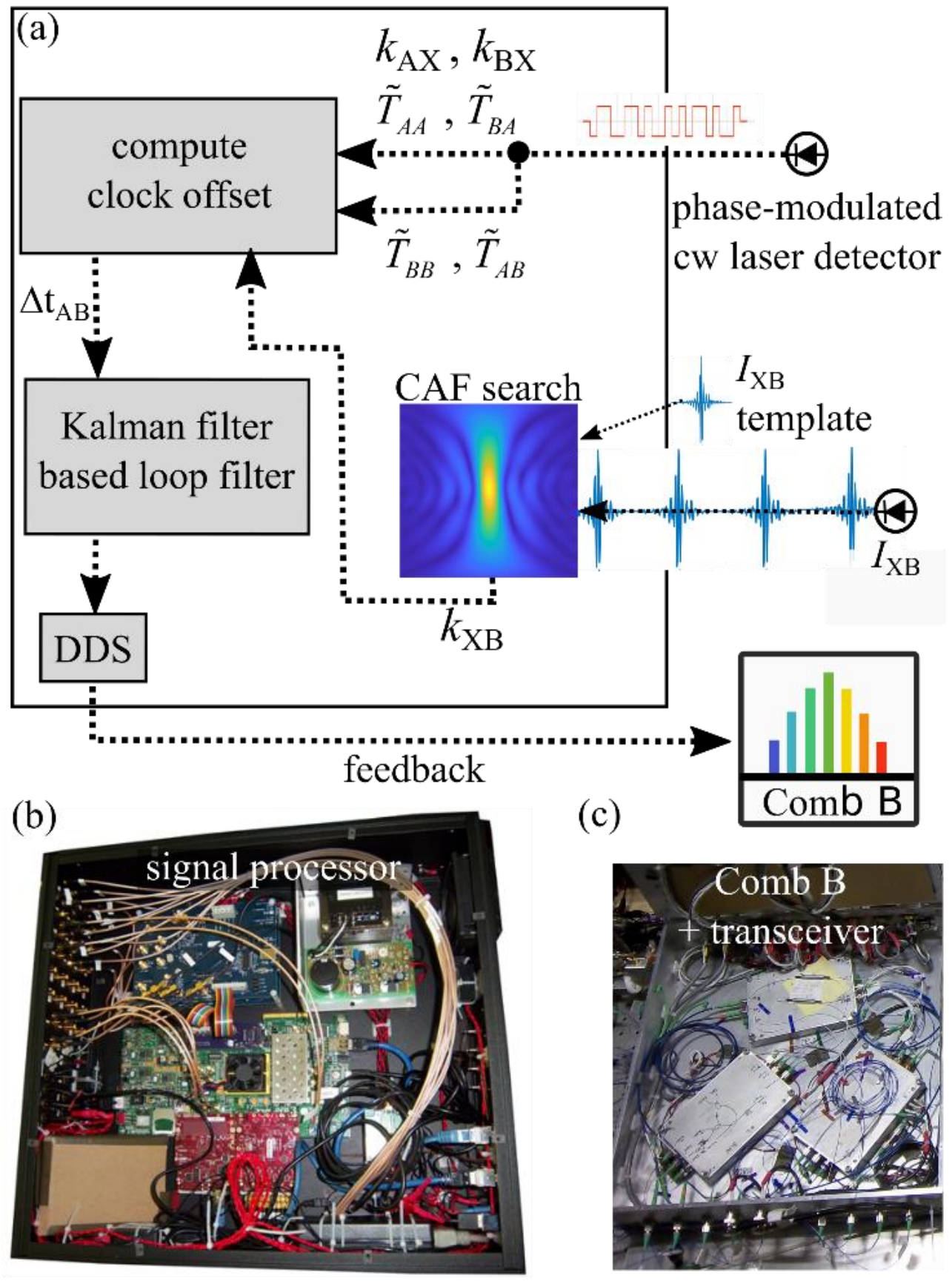

Figure 9: (a) Signal processing on Site B. The heterodyne detection between the incoming comb X and local comb B yields the series of interferograms (blue trace). After digitization and filtering, the signal processor extracts the timestamps using a cross-ambiguity function (CAF) search, as described in Appendix A. The coherent communication channel detects the heterodyne signal between the local cw laser and transmitted cw laser. As discussed in Ref. [22], it generates both the timestamps for the communication-based O-TWTFT, labelled $\{\tilde{T}_{AA}, \tilde{T}_{AB}, \tilde{T}_{BB}, \tilde{T}_{BA}\}$, and transmits the timestamps recorded on site A via data packets to site B. The signal processor computes the clock offset from the equations given in Section III. A Kalman-filter-based loop filter then applies the necessary feedback to comb B via a Direct Digital Synthesizer (DDS). (b) Image of the signal processor that uses an 8-core digital signal processor (DSP) and field programmable gate array (FPGA). (c) Image of the optical transceiver at site B. CAF: cross-ambiguity function; IGM: interferogram

The DSP determines the precise timestamp, according to the local ADC clock, corresponding to each interferogram peak. For the interferograms between comb A and X, we use a simple matched-filter-based extraction of the timestamp, as discussed in Figure 6. For the interferograms between comb B and X, we use the more complicated cross-ambiguity function (CAF) search as discussed in Appendix A. To accomplish this search in realtime, we implement first a coarse search based on an FFT algorithm along a grid defined by a few initial velocity guesses and then a fine search based on the Nelder–Mead (downhill simplex) method [30].

The detection of an interferogram also triggers the protocol to initiate a communication-based TWTFT measurement and the subsequent transmission of those data and the comb timing data from site A to site B. Thus, the communication-based TWTFT data are also updated at a rate of $\Delta f_r$. Each site takes turns at sending their own PRBS signal to generate these timestamps, $\left\{\tilde{T}_{AA}, \tilde{T}_{AB}, \tilde{T}_{BB}, \tilde{T}_{BA}\right\}$, as discussed in Appendix B and Ref. [22]., after which the communications link transmits the four timestamps from site A, $\tilde{T}_{AA}, \tilde{T}_{BA}$ , $k_{BX}$ and $k_{AX}$ to site B.

The processor on the remote site then aggregates all the observations into short circular buffers in order to handle the asynchronous data streams, allow interpolation operations (e.g. Eqn. (12)), and ensure that all observations are recent enough to calculate a clock offset. The timestamps are all offset by the calibrated transceiver delays, as discussed in Section III.B.3, in order to relate all observations to a common reference point at each site. While each observation stream is nominally sampled at a rate of $\Delta f_r \approx 2\text{kHz}$, random signal fades lead to missing timestamps. However, a single clock time offset requires a full set of timestamps from both the heterodyne comb measurement and communication-based O-TWTFT. Furthermore, the velocity estimation (Eq. (18)) requires three nearly consecutive data points. Therefore, before calculation of a time offset, we verify that all data is within a 2 ms time window. The calculated clock offset is then passed to the Kalman filter, whose phase estimate is passed to a proportional-integral loop filter that adjusts the remote comb.

## IV.C Free-space optical link, Quadcopter, and Doppler simulator

In the absence of a moving clock, we employed two different methods of generating a time-varying link both of which mimic time-transfer via moving, intermediate clock site. (See Figure 1).

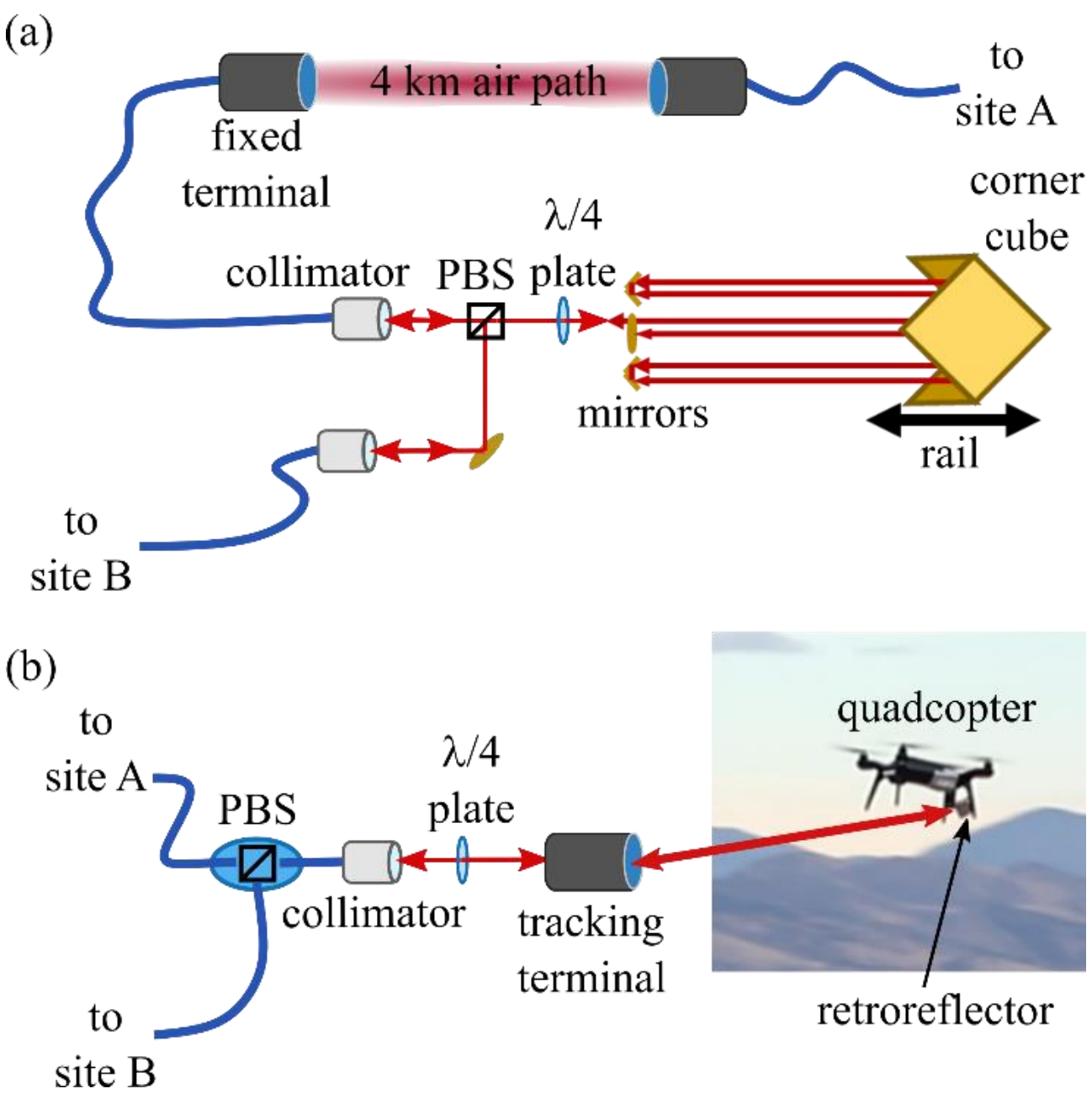

Figure 10: (a) Schematic of the Doppler-simulator. The corner cube is on a 2-meter long rail. This 6-pass design allows for bi-directional propagation, effective displacements of 24 m and effective speeds of 25 m/s. (b) Detailed schematic of coupling to quadcopter-mounted-retroreflector.

The Doppler simulator rail is shown in Fig. 10a and consists of a retroreflector mounted on a cart that travels back and forth across a 2-m long rail by a belt-and-pulley and programmable servo motor. As shown, the signals are polarization multiplexed to allow for bi-directional operation and to multiply the path length by 12, giving effective displacements of 24 meters and closing velocities of up to ±24 m/s. The motion was programmed for approximate constant velocity, other than the brief deceleration/acceleration at the turn-around points. The inevitable cross-talk between polarizations causes spurious back reflections. However, these back reflections can usually be rejected in the time domain given the pulsed nature of the signals via a windowing operation. The Doppler simulator rail is used in series with the existing 0-4 km free-space link.

The Doppler simulator allowed for repeated testing but had limited displacement. For more realistic conditions, we mounted a retroreflector onto a quadcopter, which was flown directly towards and away from the clocks at effective closing velocities of up to ±20 m/s and over displacements of 500 meters. Flight restrictions at the test site prevented longer flight paths. In previous static demonstrations (as well as for the Doppler simulator), we had created a folded free-space link by placing a flat mirror at the far point. Here, we effectively replace that flat mirror with a retroreflector to minimize the pointing requirements on the quadcopter. As a consequence, the free-space link is folded onto itself, thus a single tracking terminal is needed, and we use polarization multiplexing of the two-way optical signals to maintain bi-directionality. In transitioning to this configuration, a new full OTDR-based calibration was not performed and thus the calibration of several transceiver delays was not as precise leading to a slight degradation in performance. As described in Ref. [31], the tracking terminal follows the quadcopter by feedback to a gimbal based on image processing from a bore-sighted CMOS camera that detects the retroreflected light of an 850-nm LED beacon.

# V. Results

## V.A Time Synchronization

Here, we present results from the full time synchronization of sites A and B. The signal processor at site B returns the computed clock offset from the O-TWTFT data, the estimated instantaneous closing velocity, and estimated instantaneous time-of-flight. For verification data, it also samples the out-of-loop clock offset continuously. We then have two different measures of the clock time offset: (i) the O-TWTFT computed time offset, which is only available in the absence of signal fades, and (ii) the verification data for the clock offset, which is available at all times including during fades and during the subsequent re-synchronization. Both are available at a 2-kHz update rate. Typically, we only show the verification data as it is an “out-of-loop” measurement as opposed to the computed O-TWTFT clock offset, which is “in-loop” in the sense it must be driven to zero by the overall synchronization feedback. Figure 11a shows one-second of “out-of-loop” verification data for the clock offset when the system is operated with the Doppler simulator at effective ±24 m/s relative velocities in series with the 4-km turbulent air path. The feedback to the remote clock at site B was set to a synchronization bandwidth of 10 Hz, giving the clear shoulder in the timing power spectral density (PSD) of Fig. 11b.

Because the link included a 4-km turbulent air path, there are also signal fades of a few millisecond duration which there is no O-TWTFT data. In figure 11a, we indicate these periods by the blue data points. Since they are short, the use of the Kalman filter effectively maintains synchronization. In general, we distinguish between periods of active synchronization, when there is valid O-TWTFT data over the link, and periods without active synchronization, when there has been a long signal fade and therefore no O-TWTFT data. For the typically short signal fades, such as those evident in Figure 11a, we still consider the system actively synchronized. However, if the signal-fade is of long duration (with respect to the inverse overall feedback bandwidth) then we consider the remote site free-running. Given the current bandwidths, we consider the system “actively synchronized” if the signal fades are less than 20-ms in duration. Therefore, when we report time deviation or other quantities during active synchronization, we mask out the period of the signal fade and re-acquisition for fades greater than 20-ms duration.

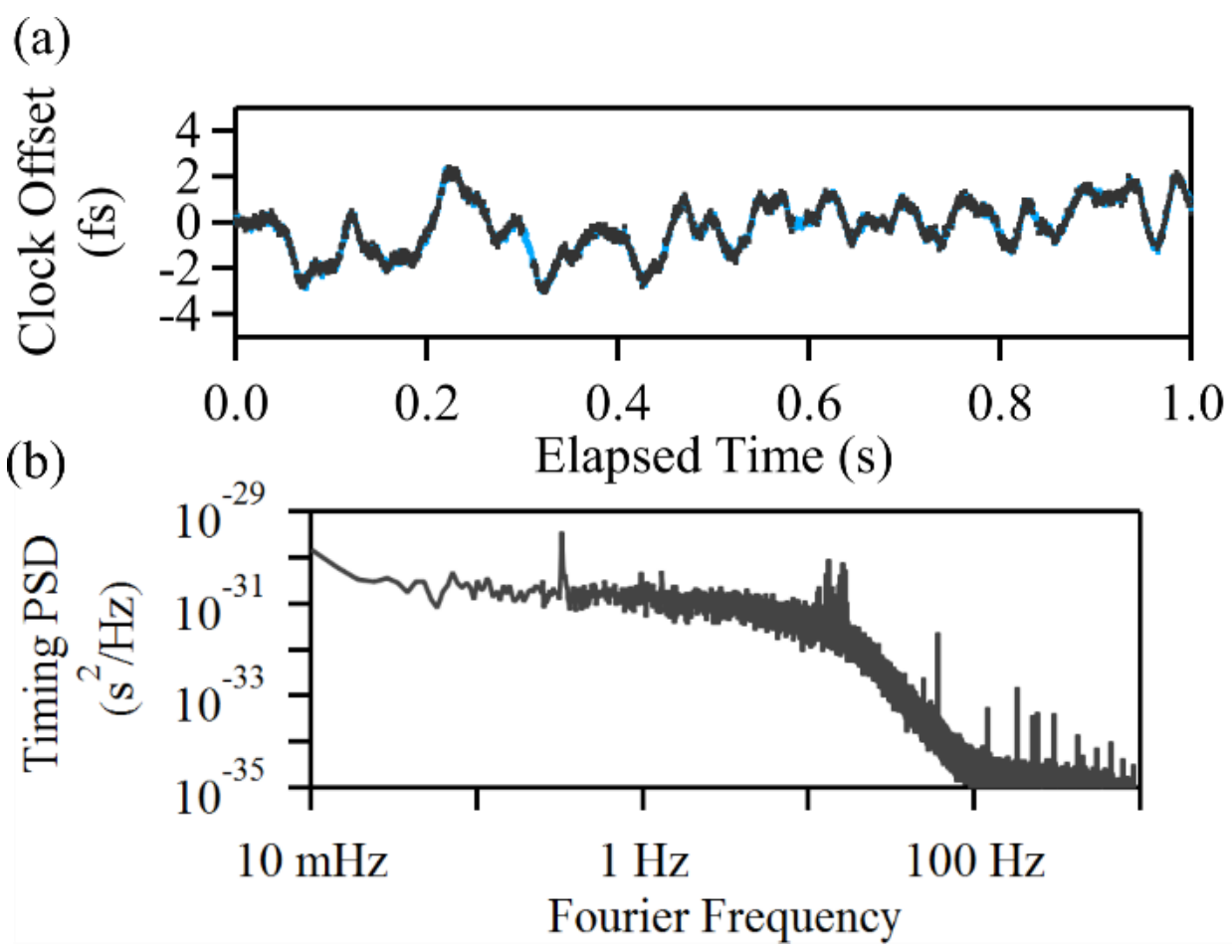

Figure 11: (a) Out-of-loop clock offset sampled continuously (light blue) and only during active synchronization (grey dots) at the full ~ 2 kHz sampling rate over a 4-km open air path and at a 24 m/s effective speed by use of the Doppler simulator. (c) Timing power spectral density (PSD) from the data in (a).

Figure 12a presents a longer 10-s data set acquired with the Doppler simulator in series with the 4-km turbulent open-air path. As expected, at the full 2 kHz sampling rate, the in-loop computed O-TWTFT clock offset is noisier than the out-of-loop verification data because the optimized feedback bandwidth of 10 Hz effectively smooths the clock offset. (In other words, for timescales shorter than 0.1 s, the timing follows the local optical oscillator.) However, its average value is zero since the system is phase-locked. Because of turbulence across the 4-km air path, there are many fades during even this brief data set, indicated by the blue regions in the out-of-loop clock offset. For several of the longer duration fades, the out-of-loop clock offset shows both the random, slow walk-off in the clock time offset followed by re-synchronization. The velocity estimate has an uncertainty of 1.2 mm/sec for the full rate data (2 kHz) due to the white noise on the timestamps from which it is computed. This uncertainty drops to 250 μm/sec in just tens of milliseconds of averaging.

Figure 12b presents the out-of-loop clock offset for operation with the Doppler simulator at ±24 m/s effective closing velocities over varying open-air paths of 0, 2, and 4 km for a 1200-s duration measurement. In addition, it presents data for zero velocity over a shorted (0-km) path. The standard deviations are 0.98 fs, 1.0 fs, 1.2 fs, and 0.81 fs, respectively. The slightly increased standard deviations are consistent with, and attributed to, an increased number of signal fades over the longer air paths rather than systematic velocity-dependent effects. The slow wander in the out-of-loop time offset evident in these data is dominated by temperature-induced variations in the path length of the optical fiber connecting the two sites to provide the verification time offset signal.

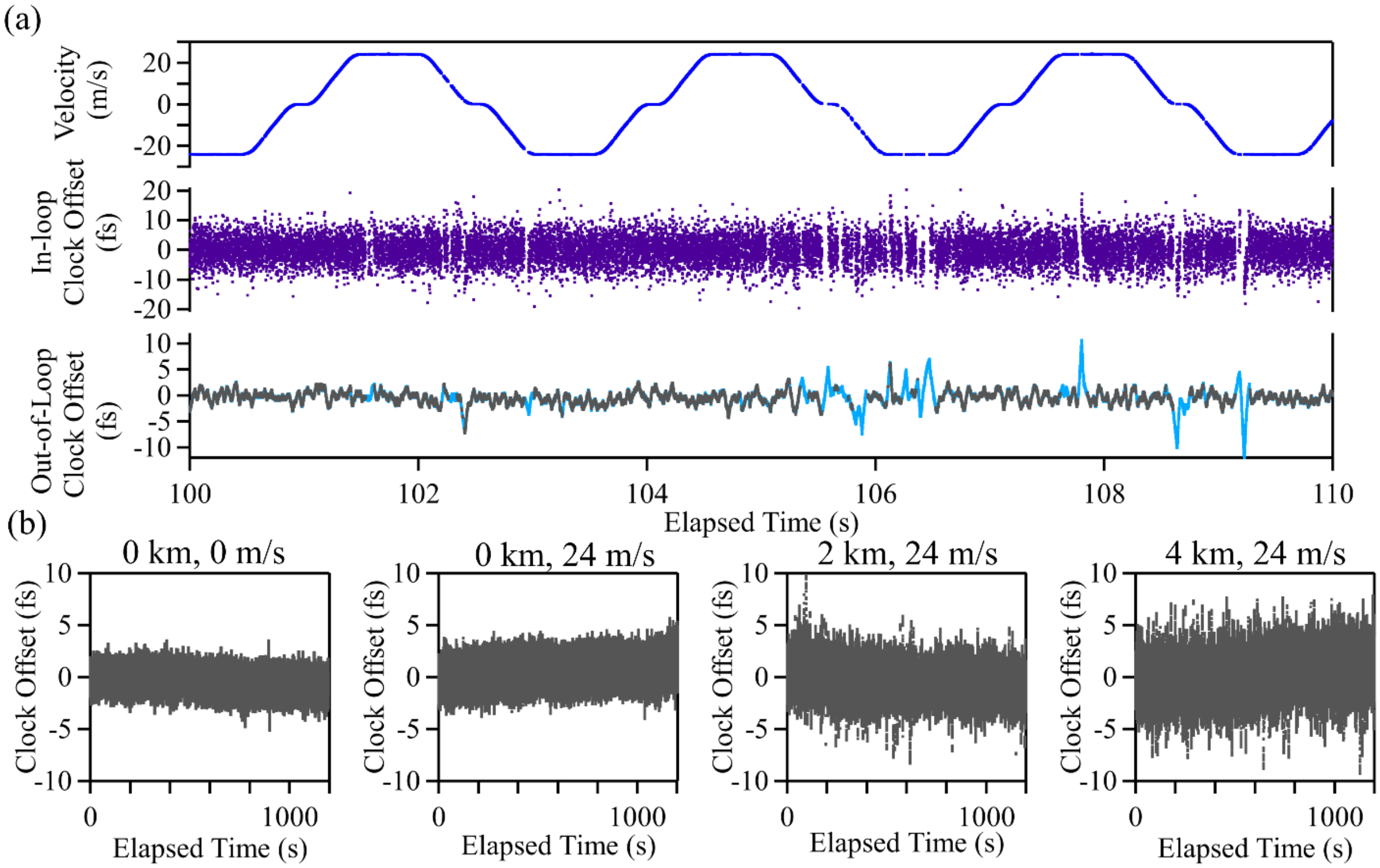

Figure 12: (a) Clock synchronization with the Doppler simulator operated at its maximum ±24 m/s speed. The plot shows the instantaneous closing velocity (dark blue) as the retroreflector cycles back and forth on the Doppler simulator rail, the in-loop computed O-TWTFT clock offset (purple dots) and the out-of-loop verification clock offset during periods of active synchronization (grey dots) and across signal fades (light blue line). (b) Out-of-loop measured clock offset during periods of active synchronization for the following conditions: $V = 0$ m/s and $L = 0$ km, $V = \pm 24$ m/s and $L = 0$ km, $V = \pm 24$ m/s and $L = 2$ km, $V = \pm 24$ m/s and $L = 4$ km. All data is at the full 2 kHz sampling rate.

Figure 13 shows a similar data set but for three passes of the quadcopter-mounted retroreflector The signal fades here can sometimes have long duration due to the challenges of coupling light into the single-mode fiber while tracking the moving quadcopter, rather than the effects of atmospheric turbulence. The standard deviation of the clock offset during active synchronization is 3.7 fs representing a slight degradation in performance from that of the Doppler simulator. We attribute this degradation to the increased measurement noise due to lower return powers and frequent signal fades, as well as the less precise delay calibration. Figure 13 includes expanded views of the clock offset during periods of active synchronization but that include short signal fades. The significant control effort to re-acquire synchronization after a long duration fades is clearly evident in the lower rightmost panel, which exhibits ~ 100 fs clock excursion.

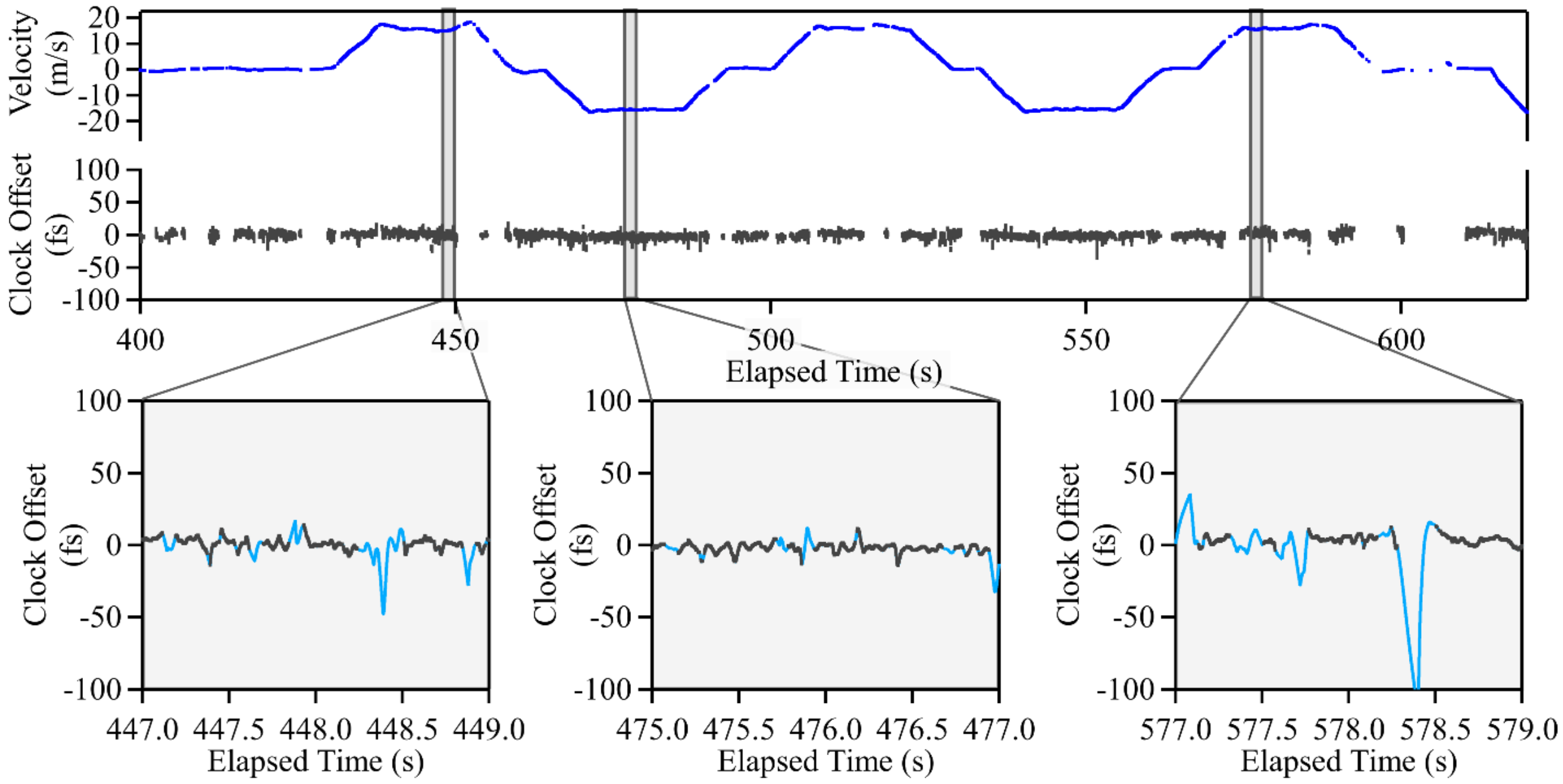

Figure 13: Clock synchronization during flight of quadcopter. Top Panel: Three passes of the quadcopter showing the instantaneous closing velocity (blue) and out-of-loop clock offset during active synchronization (grey dots). Bottom Panels: Expanded views containing continuously sampled clock offset (light blue) as well as only during active synchronization (grey dots) showing clock walk-off during signal fades and synchronization re-acquisition.

Ref. [16] provides a statistical analysis of these results in terms of a Time Deviation and Modified Allan Deviation for both the Doppler simulator and quadcopter.

## V.B Velocity-Dependent Bias

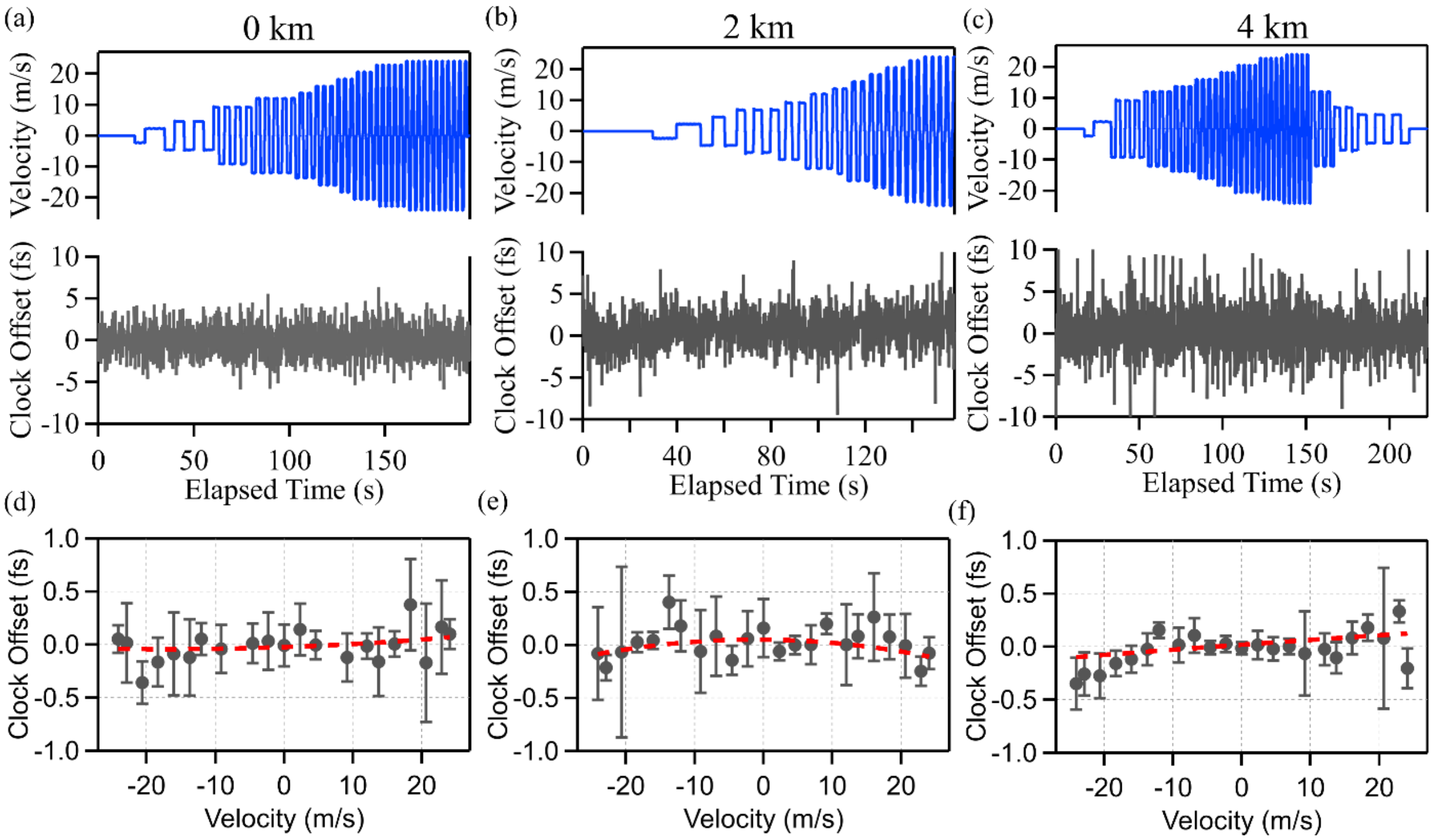

Figure 14: Time series of closing velocity and clock offset for operation of the Doppler simulator in series with a (a) 0 km, (b) 2 km, and (c) 4 km open air path. Here, to evaluate any residual velocity-dependent bias independent of feedback dynamics associated with signal fades, the clocks are phase-locked via the verification data and the reported clock offset is computed via the O-TWTFT signals. The clock offset data is resampled to 10 Hz. (d-e) Clock offset vs closing velocity (grey circles) for the 0 km, 2 km, and 4 km paths of (a)-(c). The red dashed line is a quadratic fit to the data which is used to estimate any residual bias.

Figure 14 summarizes an analysis to quantify any velocity-dependent bias based on data acquired with the Doppler simulator at effective relative velocities from 0 to ±24 m/s and a free-space path length of 0, 2 or 4 km. As discussed below, we find no bias to within 330 as uncertainty for relative velocities up to ±24 m/s.

For these data, we reversed the normal operation of the system. We re-routed the signals so that the input to the feedback for the remote clock was the verification clock offset data rather than the clock offset computed from the O-TWTFT data. In this way, we avoid the increase in the clock offset noise caused by the combination of signal fades and loop dynamics. (We could also have simply locked both sites to the same local oscillator and disabled the feedback bandwidth.) This configuration allows us to examine the clock offset computed from the O-TWTFT data for any residual velocity-dependent bias without the impact of the loop dynamics. The data of Figure 14a exhibit a slow, few-femtosecond wander, but this wander is not correlated with velocity. It is attributed to laboratory temperature fluctuations that modulate the length of the fiber optic in the verification path. As seen earlier, we do observe an increase in the clock-offset noise for the 2 and 4 km path, but again there is no correlation with velocity. It is attributed to reduction in received power due to atmospheric turbulence.

To quantify any residual velocity-dependent bias, we calculated the mean of the clock offset versus closing velocity by averaging the clock offset over periods of constant velocity from the Doppler simulator. These data are given Fig. 14 (d)-(f) for 0, 2, and 4 km free-space paths, respectively. We then fit these data to the quadratic function $\Delta t = c_0 + c_1 V + c_2 V^2$, yielding the red dashed lines of Fig. 14. In the absence of bias, the coefficients $c_1$ and $c_2$ should be zero. The values returned from these fits are given in Table II. The 'worst' case (2 sigma) linear systematic would therefore be 210 as and the 'worst' case (2 sigma) quadratic systematic would therefore be 330 as. Note that a velocity-dependent bias below 330 attoseconds represents orders of magnitude suppression of potential sources of error as illustrated in Table I.

Table II: Coefficients from weighted quadratic fit of clock offset versus closing velocity.

| Pathlength [km] | $c_0$ [as] | $c_1$ [as/(m/s)] | $c_2$ [as/(m/s)$^2$] |
|---|---|---|---|
| 0 | 19 ± 30 | 4.5 ± 2.1 | -0.01 ± 0.14 |
| 2 | 51 ± 45 | -0.5 ± 2.2 | -0.26 ± 0.16 |
| 4 | -23 ± 65 | 2.3 ± 2.7 | 0.07 ± 0.21 |

## VI. Considerations in Scaling to Higher Velocity and Longer Distances

There is no clear velocity-dependent limit in either the data of Fig. 12, 13 and 14 or the data of Ref. [16]. Rather, the short-term noise is due to the timing jitter of the frequency combs and the long term wander is due to temperature drifts in the out-of-loop verification fibers. For these data, the Doppler simulator reached a maximum speed of 24 m/s and a folded physical displacement of 12 m, while the quadcopter reached a maximum speed of 20 m/s and a folded physical displacement of 500 m. At even higher velocities or longer displacements, we expect limitations on the synchronization performance to arise due to terms of order $V^2$, the presence of accelerations, the limits of the transceiver calibration, and the configuration of the initial detection of an interferogram. A number of these effects can be calculated and the equations of Section III extended.

For example, we can consider the impact of acceleration. While the limits due to acceleration were not fully explored experimentally due to physical limitations of the Doppler simulator, the system reached accelerations of ~70 m/s$^2$. Naively, we might expect the first order contribution to scale as $\frac{a}{c}\left[\frac{1}{2\Delta f_r}+T_{\mathrm{A}\to\mathrm{B}}\right]^2$, where $a$ is the acceleration, yielding 15 fs for $a \sim 70$ m/s$^2$. However, by using centered derivatives to estimate the closing velocity at both sites at the correct time, the contribution from acceleration is below 100 attoseconds. This limit could conceivably be lowered even more if resampling of the detected phases was implemented, or an explicit model of the displacement vs time which includes an acceleration term was added to the equations.

Assuming that the higher-order velocity and acceleration terms could be handled by extending the O-TWTFT equations of section III, the most significant concern at higher velocities will be the increased Doppler shifts. The interferograms are currently sampled at $f_r$ by the local comb's pulses and afterwards by the local ADC with exactly the same $f_r$ clock. To avoid aliasing, we thus currently require the Doppler shifts to be below roughly half of our Nyquist range of 100 MHz, which limits us to an effective speed of 75 m/s. In principle, by exploiting aliasing the system ithout could support higher Doppler shift but this is restrictive in practice. Moreover, strong Doppler shifts will cause problems with the initial detection, or triggering, of the interferograms. Finally, the cross-ambiguity search for the interferogram peaks, described in Appendix A, is only suitable for a limited range of Doppler shifts in its current implementation. Many of these concerns apply as well to the coherent communication channel as the optical carrier is also frequency-shifted and may lead to the loss of interference signals. Therefore, as the speeds exceed ~50 m/s, the system will require modified approaches, such as IQ detection, proper bandpass sampling of the interferograms, separate transmit and local oscillators for the communication channel, digital filter banks for triggering etc.. However, these technical challenges are not dissimilar from those encountered in radar and coherent optical communication and a similar set of tools can be successfully applied.

# VII. Conclusion

Here, we have presented a detailed discussion of the derivation of a master synchronization equation and the digital signal processing necessary to synchronize clocks using comb-based O-TWTFT in the presence of motion. Using this implementation, we have shown that clocks at distant sites can be synchronized to below a femtosecond despite a time-varying link with associated speeds of up to ±24 m/s – a factor of 10,000 suppression of potential velocity-dependent effects. Additionally, we find no velocity-dependent bias between the synchronized clocks to within the statistical uncertainty of 330 attoseconds. This approach should scale to higher velocities as long as the considerations of calibration and Nyquist limitations of Section VI are handled appropriately. This approach opens the door for free-space clock networks between mobile, airborne or spaceborne platforms.

**Acknowledgments**

This work was funded by the Defense Advanced Research Projects Agency (DARPA) PULSE program and the National Institute for Standards and Technology (NIST) . We thank Prem Kumar, Martha Bodine, Jennifer Ellis and Kyle Beloy for helpful discussions.

# Appendix A: Compensation of Systematics Due to Delay-Doppler Coupling

Here we derive the equations governing the coupling of dispersion and Doppler effects that yield systematic time shifts. We consider second-order dispersion only, but the equations can be generalized to higher order dispersion. Note that the algorithm uses a fit to the cross-ambiguity function, which does correct for these higher order dispersion effects.

Consider the interferogram $I_{\mathrm{BX}}$. As a function of oracle time and including the response of the rf detection system, it is given by

$$\begin{aligned} I_{\mathrm{BX}}\left(t, z_{\mathrm{A}}\right) &= h_{\mathrm{rf}}\left(t\right) * \left(E_{\mathrm{B}}^{*}\left(t, z_{\mathrm{A}}\right) E_{X}\left(t, z_{\mathrm{A}}\right)\right) \\ &\approx h_{\mathrm{rf}}\left(t\right) * \left[\exp\left(i2\pi\left(\Delta\tilde{\nu}_{\mathrm{XB}} + \nu_{\mathrm{Doppler}}\right)t\right) \sum_{m} E_{\mathrm{B},m}^{*} E_{\mathrm{X},m} \exp\left(i2\pi m \Delta f_{r}\, t + i2\pi m f_{r}\, \tau\right)\right] \end{aligned} \tag{22}$$

where $\Delta\tilde{\nu}_{\mathrm{XB}} = \tilde{\nu}_{\mathrm{X}} - \tilde{\nu}_{\mathrm{B}}$, $\tilde{\nu}_{\mathrm{X}}$ is the frequency of some central tooth of comb X, $\nu_{\mathrm{Doppler}} \equiv \left(V/c\right)\tilde{\nu}_{\mathrm{B}}$ is comb B's doppler shift, $*$ is the convolution operator, and $h_{\mathrm{rf}}(t)$ models the lumped rf detection chain's impulse response. We dropped any constant phase terms and simplified the phase difference to $\Phi_{\mathrm{X}}(t, z_{\mathrm{A}}) - \Phi_{\mathrm{B}}(t-\tau, z_{\mathrm{A}}) = 2\pi m \Delta f_r t + 2\pi m f_r \tau$, where $\tau$ is the desired time stamp equal to the time-of-flight for comb B modulo $f_r$. The objective is to measure $\tau$ independent of $\nu_{\mathrm{Doppler}}$. In writing (22), we make several approximations. First, we assume this detector response has low enough bandwidth that we keep only a single term of the double-sum as in Eqn. (5). Second, we apply a common Doppler shift to all comb teeth, corresponding to the Doppler shift on the tooth at the center of the transmitted comb B. Here, we neglect the corresponding Doppler shift on the repetition rate. The effect of this Doppler shift on the spacing of successive timestamps is, of course, included within the equations of Section III, but here we are concerned only with the time shift of a single interferogram. In this case, the Doppler shift on the repetition frequency only causes a negligible stretching of the interferogram. (It would have to be included at higher velocities or if $\Delta f_r$ were significantly lower.)

Consider Eq. (22) in the frequency domain at zero velocity and zero Doppler shift. For notational simplicity, we drop the BX subscript and the z-argument to find,

$$\begin{aligned} I(f) &\equiv \mathcal{F}\left\{I_{\mathrm{BX}}(t)\right\} \\ &\equiv H_{\mathrm{rf}}(f)\mathcal{F}\left\{\exp\left(i2\pi\left(\Delta\tilde{\nu}_{\mathrm{XB}}+\nu_{\mathrm{Doppler}}\right)t\right)\sum_m A_{\mathrm{B}}^{*}(mf_r)A_{\mathrm{X}}(mf_r+m\Delta f_r)\exp\left(i2\pi m\Delta f_r t+i2\pi m f_r\tau\right)\right\} \\ &= S\left(f,\tau,\nu_{\mathrm{Doppler}}\right)\sum_m \delta\left(f-m\Delta f_r-\Delta\tilde{\nu}_{\mathrm{XB}}+\nu_{\mathrm{Doppler}}\right) \end{aligned} \quad (23)$$

where we define the baseband comb spectral envelopes $A_{\mathrm{X}}(f)\equiv \mathrm{E}_{\mathrm{X}}\left(f+\tilde{\nu}_{\mathrm{X}}\right)$ and $A_{\mathrm{B}}(f)\equiv \mathrm{E}_{\mathrm{B}}\left(f+\tilde{\nu}_{\mathrm{B}}\right)$, and

$$\begin{aligned} &S\left(f,\tau,\nu_{\mathrm{Doppler}}\right) \\ &= H_{\mathrm{rf}}(f)A_{\mathrm{B}}^{*}\left(M\left[f-\Delta\tilde{\nu}_{\mathrm{XB}}+\nu_{\mathrm{Doppler}}\right]\right)A_{\mathrm{X}}\left((M+1)\left[f-\Delta\tilde{\nu}_{\mathrm{XB}}+\nu_{\mathrm{Doppler}}\right]\right)\exp\left(i2\pi M\left[f-\Delta\tilde{\nu}_{\mathrm{XB}}+\nu_{\mathrm{Doppler}}\right]\tau\right) \end{aligned}$$

(24)

where $M\equiv f_r/\Delta f_r$ is the "time expansion" or "frequency contraction" provided by the linear optical sampling. The inverse Fourier transform of Eq. (23) is simply a train of interferograms, where each individual interferogram has a spectrum $S\left(f,\nu_{\mathrm{Doppler}}\right)$. Note that in the case where the frequency shifts are negligible, $M>>1$, and there is no time delay,

$$S(f,0,0)\approx H_{\mathrm{rf}}(f)A_{\mathrm{B}}^{*}(Mf)A_{\mathrm{X}}(Mf) \quad (25)$$

which is just the product of the scaled comb pulse spectra, modified by the rf detector response.

To find the time delay, we consider the spectral phase of (24)

$$\angle S\left(f,\tau,\nu_{\text{Doppler}}\right)\approx 2\pi Mf\tau+\pi^2\beta_{2,\text{rf}}f^2+\pi^2\Delta\beta_2 M^2\left[f-\Delta\tilde{\nu}_{\text{XB}}+\nu_{\text{Doppler}}\right]^2 \tag{26}$$

where we assume the group delay associated with the rf detection chain is included in calibration and drop higher-order dispersion terms. The first term yields the desired time delay and the last two are quadratic variations in the spectral phase from second-order dispersion that can lead to systematic errors.

To remove the dispersion terms and therefore temporally narrow the interferogram which improves the SNR, we can apply a matched filter or simply multiply by $S^*(f,0,0)$, measured during calibration. We then have the simple relationship,

$$\angle\left\{S\left(f,\tau,0\right)S^*\left(f,0,0\right)\right\}=2\pi Mf\tau\ , \tag{27}$$

in the absence of Doppler shifts. In the real-time processor, we actually find the peak of the filtered time-domain signal but the result is the same and avoids problematic spectral phase unwrapping at low SNR. Now consider the bias in the presence of Doppler shifts, where

$$\begin{aligned}\angle\left\{S\left(f,\tau,\nu_{Doppler}\right)S^*\left(f,0,0\right)\right\}&=2\pi Mf\tau+\pi^2\Delta\beta_2 M^2\left[f-\Delta\tilde{\nu}_{XB}\right]^2-\pi^2\Delta\beta_2 M^2\left[f-\Delta\tilde{\nu}_{XB}+\nu_{Doppler}\right]^2\\&=2\pi Mf\left[\tau-\pi\Delta\beta_2 M\nu_{Doppler}\right]-2\pi^2\Delta\beta_2 M^2\nu_{Doppler}\Delta\nu_{XB}\end{aligned} \tag{28}$$

which would give a systematic velocity-dependent bias in the extracted optical pulse delay of $-\pi\Delta\beta_2 M\nu_{\text{Doppler}}$.

We remove this systematic in two steps. Fortunately, the rf dispersion term of Eq. (26) is independent of the Doppler shift and can thus be compensated for by simply applying an inverse filter $H_{rf}^{-1}(f)$. Second, we search for the optimal Doppler shift that flattens the spectral phase and therefore leads to the strongest time-domain signal. We conduct this search in two-dimensions to both retrieve the optimal Doppler shift and the corresponding time delay. This amounts to maximizing the cross-ambiguity function by computing $I_{\mathrm{CAF}}(\tau,\nu)=\int I_{\mathrm{BX}}(t)\,I_{\mathrm{template}}(t-\tau)\exp\left(i2\pi\nu[t-\tau]\right)dt$ where $I_{\mathrm{template}}(t)=\mathcal{F}^{-1}\left\{S(f,0,0)\right\}$. Figure 8 shows an example of the computation of $I_{\mathrm{CAF}}(\tau,\nu)$ for the system with and without DCF.

# Appendix B: Ambiguity resolution by the communication-channel O-TWTFT

As discussed after Eqns. (6)-(11), we must resolve the ambiguity associated with the comb's pulse train, i.e. $p_{\mathrm{BX}}$, and $p_{\mathrm{XB}}$. (Recall that $p_{\mathrm{AX}}$ is determined from the straightforward tracking of the number of interferogram peaks between the co-located master and transfer combs.)

We perform this resolution via the communications-based TWTFT which runs in parallel on the same free-space link and produces pairs of observations of the phase of the master and remote combs at both sites. These observations have two salient characteristics: first, they are completely unambiguous and second, since they originate from a direct-modulation time transfer link, they are much less precise and accurate than the observations produced by the interferometric comb subsystem. (See Ref. [22] for details of the comm-based O-TWTFT system.)

These two characteristics make these 'coarse' (10's of ps) but unambiguous observations perfectly suited to resolve the ambiguities present in the comb equations. Any noise on the coarse observations will drop out of the final result, as long as it is less than half a period of the ambiguities (5 ns), because it will not change the resolved integers. We note that for a system operated at a greatly increased repetition frequency, the precision of the unambiguous observations must correspondingly increase.

This communication-based O-TWTFT system generates its own set of two-way values based on the emission and arrival of a pseudorandom binary sequence: $\left\{\tilde{T}_{\mathrm{AA}}, \tilde{T}_{\mathrm{AB}}, \tilde{T}_{\mathrm{BB}}, \tilde{T}_{\mathrm{BA}}\right\}$, where the tilde indicates information associated with the lower precision comm-based O-TWTFT subsystem These four measurements contain the same, albeit noisier, information as Eqn. (15)a-d but measured at different oracle times $t_{\mathrm{AB}}$ and $t_{\mathrm{BA}}$ instead of $t_{\mathrm{XB}}$ and $t_{\mathrm{BX}}$; however, we can interpolate these values to the times $t_{\mathrm{XB}}$ and $t_{\mathrm{BX}}$. We essentially invert the equations analogous to Eqn. (13) and Eqn. (14) to find $\tilde{\Phi}_{\mathrm{A}}\left(t_{\mathrm{AB}}-T_{\mathrm{A}\to\mathrm{B}}\left(t_{\mathrm{AB}}\right), z_{\mathrm{B}}\right)$, $\tilde{\Phi}_{\mathrm{B}}\left(t_{\mathrm{AB}}, z_{\mathrm{B}}\right)$, $\tilde{\Phi}_{\mathrm{A}}\left(t_{\mathrm{BA}}, z_{\mathrm{A}}\right)$ and $\tilde{\Phi}_{\mathrm{B}}\left(t_{\mathrm{BA}}-T_{\mathrm{B}\to\mathrm{A}}\left(t_{\mathrm{BA}}\right), z_{\mathrm{A}}\right)$, where again the tilde indicates information associated with the comm-based O-TWTFT subsystem. To find the integer $p_{\mathrm{BX}}$, we use Eqn. (8), the identity $\Phi_{\mathrm{B}}\left(t_{\mathrm{pBX}}-T_{\mathrm{B}\to\mathrm{A}}\left(t_{\mathrm{pBX}}\right), z_{\mathrm{B}}\right) \equiv \Phi_{\mathrm{B}}\left(t_{\mathrm{pBX}}, z_{\mathrm{A}}\right)$, and the local $\Phi_{\mathrm{X}}\left(t_{\mathrm{pBX}}, z_{\mathrm{A}}\right)$ from the comb-based measurements to find its value through:

$$p_{\mathrm{BX}} = \mathrm{round}\left[\Phi_{\mathrm{X}}\left(t_{\mathrm{pBX}}, z_{\mathrm{A}}\right)/2\pi - \tilde{\Phi}_{\mathrm{B}}\left(t_{\mathrm{pBX}}-T_{\mathrm{B}\to\mathrm{A}}\left(t_{\mathrm{pBX}}\right), z_{\mathrm{B}}\right)/2\pi\right]. \quad (29)$$

Likewise, to resolve the integer $p_{XB}$, we first isolate it by Eqn. (10) and the identity $\Phi_{\mathrm{X}}\left(t_{\mathrm{pXB}}-T_{\mathrm{A}\to\mathrm{B}}\left(t_{\mathrm{pXB}}\right), z_{\mathrm{A}}\right) \equiv \Phi_{\mathrm{X}}\left(t_{\mathrm{pXB}}, z_{\mathrm{B}}\right)$. We then use the extrapolation expression of Eqn.(12) to generate values of $\tilde{\Phi}_{\mathrm{A}}$ at the necessary times along with the local $\Phi_{\mathrm{B}}\left(t_{\mathrm{pXB}}, z_{\mathrm{B}}\right)$ from the comb-based measurements, to find:

$$\begin{aligned} p_{\mathrm{XB}} = \mathrm{round}\Big[&\tilde{\Phi}_{\mathrm{A}}\left(t_{\mathrm{pXB}}-T_{\mathrm{A}\to\mathrm{B}}\left(t_{\mathrm{pXB}}\right), z_{\mathrm{A}}\right)/2\pi \\ &+\tfrac{\Delta f_r}{\hat{f}_r}\left\{\tilde{\Phi}_{\mathrm{A}}\left(t_{\mathrm{pXB}}-T_{\mathrm{A}\to\mathrm{B}}\left(t_{\mathrm{pXB}}\right), z_{\mathrm{A}}\right)/2\pi - \tilde{\Phi}_{\mathrm{A}}\left(t_{\mathrm{pAX}}, z_{\mathrm{A}}\right)/2\pi\right\} \\ &-\Phi_{\mathrm{B}}\left(t_{\mathrm{pXB}}, z_{\mathrm{B}}\right)/2\pi + p_{\mathrm{AX}}\Big]. \end{aligned} \quad (30)$$

## Appendix C: Derivation of Effective Timestamps in terms of measured quantities.

Here, we derive Equations (15)a-d. Equations (15)b and (15)d follow directly from the definitions, (13) and (14), combined with Eqns. (11) and (9), respectively.

Equation (15)a is slightly more challenging to derive. From Eqn. (12), we have

$$\Phi_{\mathrm{A}}\left(t_{\mathrm{pXB}}-T_{\mathrm{A}\rightarrow\mathrm{B}}, z_{\mathrm{A}}\right)=\Phi_{\mathrm{X}}\left(t_{\mathrm{pXB}}-T_{\mathrm{A}\rightarrow\mathrm{B}}, z_{\mathrm{A}}\right)-2\pi p_{\mathrm{AX}}-2\pi\Delta f_r\left\{t_{\mathrm{pXB}}-T_{\mathrm{A}\rightarrow\mathrm{B}}-t_{\mathrm{pAX}}\right\}. \tag{31}$$

We replace the first term with $\Phi_{\mathrm{X}}\left(t_{\mathrm{pXB}}-T_{\mathrm{A}\rightarrow\mathrm{B}}, z_{\mathrm{A}}\right)=\Phi_{\mathrm{X}}\left(t_{\mathrm{pXB}}, z_{\mathrm{B}}\right)=2\pi p_{\mathrm{XB}}+2\pi k_{\mathrm{XB}}$ based on Eqns. (10) and (11). We convert the time interval, $\left\{t_{\mathrm{pXB}}-T_{\mathrm{A}\rightarrow\mathrm{B}}-t_{\mathrm{pAX}}\right\}$, from measurements in "oracle" time to measurements referenced to the site A timebase as,

$$t_{\mathrm{pXB}}-T_{\mathrm{A}\rightarrow\mathrm{B}}\left(t_{\mathrm{pXB}}\right)-t_{\mathrm{pAX}}=\frac{\Phi_{\mathrm{X}}\left(t_{\mathrm{pXB}}-T_{\mathrm{A}\rightarrow\mathrm{B}}\left(t_{\mathrm{pXB}}\right), z_{\mathrm{A}}\right)-\Phi_{\mathrm{X}}\left(t_{\mathrm{pAX}}, z_{\mathrm{A}}\right)}{2\pi\left(\hat{f}_r+\Delta f_r\right)}, \tag{32}$$

which can be further be expressed in terms of measured quantities as

$$t_{\mathrm{pXB}}-T_{\mathrm{A}\rightarrow\mathrm{B}}\left(t_{\mathrm{pXB}}\right)-t_{\mathrm{pAX}}=\frac{p_{\mathrm{XB}}+k_{\mathrm{pXB}}-p_{\mathrm{AX}}-k_{\mathrm{pAX}}}{\hat{f}_r+\Delta f_r}. \tag{33}$$

Substituting into (31) yields:

$$\begin{aligned}\Phi_{\mathrm{A}}\left(t_{\mathrm{pXB}}-T_{\mathrm{A}\rightarrow\mathrm{B}}\left(t_{\mathrm{pXB}}\right), z_{\mathrm{A}}\right)&=2\pi k_{\mathrm{pXB}}\\&-\frac{2\pi\Delta f_r}{\hat{f}_r+\Delta f_r}\cdot\left[p_{\mathrm{XB}}+k_{\mathrm{pXB}}-p_{\mathrm{AX}}-k_{\mathrm{pAX}}\right]-2\pi p_{\mathrm{AX}}+2\pi p_{\mathrm{XB}}\end{aligned}. \tag{34}$$

Equation (15)c is derived by noting the identity $\Phi_{\mathrm{B}}\left(t_{\mathrm{pBX}} - T_{\mathrm{B}\rightarrow\mathrm{A}}\left(t_{\mathrm{pBX}}\right), z_{\mathrm{B}}\right) \equiv \Phi_{\mathrm{B}}\left(t_{\mathrm{pBX}}, z_{\mathrm{A}}\right)$, and then solving for $\Phi_{\mathrm{B}}\left(t_{\mathrm{pBX}}, z_{\mathrm{A}}\right)$ by use of Eqns. (6) through (9), Eqn. (12) and the relationship between time and sample number $t_{\mathrm{pBX}} - t_{\mathrm{pAX}} = \hat{f}_r^{-1}\left(k_{\mathrm{pBX}} - k_{\mathrm{pAX}}\right)$.